\documentclass[
    reprint,
    twocolumn,
    amsmath,amssymb,amsfonts,
    aps, physrev,
    superscriptaddress,
    10pt,
    bibnotes,
    longbibliography
    ]{revtex4-2}

\usepackage{bm}
\usepackage[T1]{fontenc}
\usepackage{inputenc}
\usepackage{upgreek,babel,calc,mathrsfs,textgreek,bbm}
\usepackage{amsmath,amssymb,mathtools,graphicx,xcolor}
\definecolor{bluee}{rgb}{0.0, 0.53, 0.74}

\newcommand{\fpr}{\eta^+}
\newcommand{\fnr}{\eta^-}
\DeclareMathOperator*{\argmax}{arg\,max}

\usepackage[unicode=true,pdfusetitle,
bookmarks=true,bookmarksnumbered=false,bookmarksopen=false,
breaklinks=false,pdfborder={0 0 1},backref=false,colorlinks=true]
{hyperref}

\begin{document}

	\title{Small-Coupling Dynamic Cavity: \\a Bayesian mean-field framework for epidemic inference}
	
	\author{Alfredo Braunstein}
	%\email{alfredo.braunstein@polito.it}
	\affiliation{Institute of Condensed Matter Physics and Complex Systems,
		Department of Applied Science and Technology, Politecnico di Torino,
		C.so Duca degli Abruzzi 24, I-10129 Torino, Italy}
	\affiliation{Italian Institute for Genomic Medicine (IIGM) and Candiolo Cancer Institute IRCCS, str.~prov.~142, km 3.95, Candiolo (TO) I-10060, Italy}
  \affiliation{INFN, Turin Via Pietro Giuria 1, I-10125 Turin, Italy}
	
  \author{Giovanni Catania}
	\affiliation{Departamento de Física Téorica, Universidad Complutense, 28040 Madrid, Spain}
	
  \author{Luca Dall'Asta}
	\affiliation{Institute of Condensed Matter Physics and Complex Systems,
		Department of Applied Science and Technology,
		Politecnico di Torino, C.so Duca degli Abruzzi 24, I-10129 Torino, Italy}
  \affiliation{Italian Institute for Genomic Medicine (IIGM) and Candiolo Cancer Institute IRCCS, str.~prov.~142, km 3.95, Candiolo (TO) I-10060, Italy}
  \affiliation{INFN, Turin Via Pietro Giuria 1, I-10125 Turin, Italy}
	
  \author{Matteo Mariani}
	%\email{matteo.mariani@polito.it}
  \affiliation{Institute of Condensed Matter Physics and Complex Systems,
  Department of Applied Science and Technology,
  Politecnico di Torino, C.so Duca degli Abruzzi 24, I-10129 Torino, Italy}
	\affiliation{Department of Neurobiology, David Geffen School of Medicine, University of California, Los Angeles, CA 90095, USA}
	
  \author{Fabio Mazza}
	%\email{fabio.mazza@polito.it}
  \affiliation{Institute of Condensed Matter Physics and Complex Systems,
  Department of Applied Science and Technology,
  Politecnico di Torino, C.so Duca degli Abruzzi 24, I-10129 Torino, Italy}
  \affiliation{Department of Electronics, Information and Bioengineering, Politecnico di Milano, Via Ponzio 34/5, I-20133 Milano, Italy}
	
  \author{Mattia Tarabolo}
	\email{mattia.tarabolo@polito.it}
	\affiliation{Institute of Condensed Matter Physics and Complex Systems,
		Department of Applied Science and Technology,
		Politecnico di Torino, C.so Duca degli Abruzzi 24, I-10129 Torino, Italy}

	\begin{abstract}
		A novel generalized mean field approximation, called the Small-Coupling Dynamic Cavity (SCDC) method, for epidemic inference and risk assessment is presented. The method is developed within a fully Bayesian framework and accounts for non-causal effects generated by the presence of observations. It is based on a graphical model representation of the epidemic stochastic process and utilizes dynamic cavity equations to derive a set of self-consistent equations for probability marginals defined on the edges of the contact graph. By performing a small-coupling expansion, a pair of time-dependent cavity messages is obtained, which capture the probability of individual infection and the conditioning power of observations. In its efficient formulation, the computational cost per iteration of the SCDC algorithm is linear in the duration of the epidemic dynamics. While the method is derived for the Susceptible-Infected (SI) model, it is straightforwardly applicable to many other Markovian epidemic processes, including recurrent ones. This linear complexity is particularly advantageous for recurrent epidemic processes, where inference methods are typically exponentially complex in the duration of the epidemic dynamics. The method exhibits high accuracy in assessing individual risk, as demonstrated by tests on the SI model applied to various classes of synthetic contact networks, where it performs on par with Belief Propagation techniques and generally exceeds the performance of individual-based mean-field methods. Additionally, the method was applied to recurrent epidemic models, where it showed interesting performance even for relatively large values of the infection probability, highlighting its versatility and effectiveness in challenging scenarios. Although convergence issues may arise due to long-range correlations in contact graphs, the estimated marginal probabilities remain sufficiently accurate for reliable risk estimation. Future work includes extending the method to non-Markovian recurrent epidemic models and investigating the role of second-order terms in the small coupling expansion of the observation-reweighted Dynamic Cavity equations.
	\end{abstract}
	
	\maketitle
	
	\section{Introduction}\label{sec:intro}
    
	In the past decade, the increasing availability of detailed epidemiological data and high-accuracy contact-network datasets triggered the study of individual-based epidemic inference problems, which aim to predict the marginal probabilities of infection for individual nodes in a network, given partial observations of the epidemic, such as the states of specific individuals at certain times. The interest has been further stimulated during the COVID-19 pandemic, by the possibility of performing massive epidemic surveillance and digital contact tracing via smartphone applications \cite{ferretti2020quantifying,baker2021epidemic}. A variety of computational methods were proposed for tackling this class of inference problems, such as heuristic algorithms based on network centrality measures \cite{shah2010detecting,shah2011rumors}, generalized mean-field approximations \cite{lokhov2014inferring}, Monte Carlo methods \cite{antulov2015identification}, and machine learning techniques exploiting tailored architectures of autoregressive and graph neural networks \cite{biazzo2022bayesian,braunstein2023inference,shahFindingPatientZero2020a,tanDeepTraceLearningOptimize2024,9616110}. The leading techniques for epidemic inference adopts a Bayesian framework, with a simple individual-based epidemic model as prior distribution and the sparse observation of positive/negative test results as evidence \cite{ghioBayesoptimalInferenceSpreading2023a}. Unfortunately, the associated inference task includes the formidable computational difficulty of computing averages on a complicated high dimensional discrete distribution for which no tractable analytic solution is known. Arguably, generalized mean-field methods and their associated algorithms, are best able to balance the computational demand of accurately estimating these observables with affordable calculation times.
    In particular, Belief Propagation (BP) \cite{altarelli2014bayesian,altarelli2014patient,braunstein2016inference,bindi2017predicting} has proven to be extremely effective in estimating local marginal probabilities of the posterior distribution, reconstructing the infection state of unobserved individuals, and identifying patient zero and contagion channels. Furthermore, when integrated into the framework of digital contact tracing for COVID-19, the BP-based algorithms have been shown to provide a better assessment of individual risk and improve the mitigation impact of non-pharmaceutical intervention strategies, outperforming competing methods for various epidemic inference problems defined on contact networks \cite{baker2021epidemic, muntoniEffectivenessProbabilisticContact2024}. BP-based algorithms have also been used to optimally deploy resources for constraining the epidemic spreading to hit a subset of target nodes at specific times \cite{lokhovOptimalDeploymentResources2017}, similarly to the constraints imposed by observations in inference. The Belief Propagation approach to spatio-temporal epidemic trajectories can be classified as a generalized mean-field method because the epidemic trajectories of the neighbors of a given individual are assumed to be conditionally independent. This hypothesis is correct when the dynamical process takes place on a contact network without cycles, but the method proved to be very effective also on contact networks with cycles. It should be noted that the Dynamic Cavity (DC) approach \cite{neri2009cavity,kanoria2011majority} turns out to be equivalent to BP in the case of pure time-forward epidemic dynamics without observations. Moreover, for dynamic models with non-recurrent individual states, in the case of pure time-forward epidemic dynamics without observations, the BP/DC approach simplifies into a dynamic message passing technique that has been extensively used to study spreading processes on networks \cite{lokhov2015dynamic,altarelli2013large}.

    Despite the apparent success of the BP approach, simpler and faster methods may be preferable, in particular in views of the potential need of real-time calculation for large communities (with thousands to millions of individuals). Moreover, BP is not easily generalizable to recurrent dynamics (such as SIS and SIRS), in which even the trajectory of a single individual can potentially take an exponential number of states. In this regime, further assumptions are required. One possibility is a tensorial approximation of distributions of such trajectories, which can however result in relatively heavy computational requirements \cite{crottiMatrixProductBelief2023}. In this context, simpler individual-based mean-field (IBMF) methods may turn out to be of interest. Also known as quenched mean-field and N-intertwined models, these methods simply assume that the simultaneous states of neighboring nodes are statistically independent, and have been shown to provide reasonably accurate approximations to time-forward epidemic dynamics \cite{gomez2010discrete,van2008virus,kiss2017mathematics,pastor2015epidemic}. Recently, the individual-based mean-field method has been applied to propose a straightforward inference approach that heuristically incorporates observations of individual states \cite{baker2021epidemic}. While this method does not explicitly rely on a Bayesian framework, its simplicity and computational efficiency make it a practical tool for epidemic risk assessment, yielding promising results in relevant scenarios.
    
    In this paper we propose Small-Coupling Dynamic Cavity (SCDC), a novel generalized mean field approximation for Bayesian epidemic inference and risk assessment. The method is introduced to address the limitations of BP in handling recurrent epidemic processes and to ensure uniform computational complexity for both recurrent and non-recurrent models. Moreover, SCDC offers a simple and interpretable approximation of BP, where the messages have a clear physical interpretation, making it particularly appealing for a better understanding of the inference algorithm. The method exploits the improved uncorrelation assumption of BP/DC compared to simpler mean-field approaches, while largely preserving the analytic simplicity and computational efficiency of IBMF. The starting point of the method is a graphical model representation of the epidemic stochastic process that allows for a convenient derivation of a set of dynamic cavity equations for functional probability marginals defined on the edges of the contact graph. In this representation, the cavity probability marginal on a directed edge from individual $i$ to individual $j$ is a function of two quantities: the trajectory of the individual state of $i$ in the absence of interactions with $j$ and a conjugate external field acting on $i$ (which replaces the effect of the missing interaction terms in the cavity graph). By performing an expansion of the dynamic cavity equations for weak infection probabilities and truncating the expansion at the first order, a set of self-consistent equations for the average of these two quantities can be obtained. Unlike other individual-based mean-field approximations, the proposed method is conceived as an approximation of a fully Bayesian framework. While both SCDC and IBMF account for non-causal effects generated by the presence of observations, the Bayesian foundation of SCDC makes it a promising basis for future developments, such as improved approximations or theoretical studies of epidemiological inference. These non-causal effects arise because incorporating observations constrains the dynamics to trajectories compatible with the observed data, necessitating the propagation of information backward in time, which in our formulation is encoded through the conjugate fields. In the absence of observations, the conjugate fields responsible for non-causal dynamics vanish, and the individual-based mean-field method for time-forward dynamics can be recovered. For simplicity, we initially present the SCDC in the case of the Susceptible-Infected model. Using an efficient formulation based on a transfer-matrix technique, the SCDC method is then extended to general Markovian epidemic processes, including individual recovery, latency, and recurrent infection (e.g. SIR, SEIR, SIS, SIRS models). Although SCDC is expected to be less accurate than BP (on non-recurrent models on which it can be applied), for which it represents a sort of weak-infectivity approximation for the distribution of individual trajectories, we show through extensive simulations on realistic examples how this gap is often negligible.

	The manuscript is organized as follows: Section \ref{sec:methods} presents the SCDC method and its derivation on the SI model; Section \ref{sec:efficientformulation} discusses a general efficient formulation of the algorithm that is easy to generalize to other epidemic models (further discussed in Section \ref{sec:generalization_othermodels}); results are presented in Section \ref{sec:results}, concerning both estimates of epidemic outbreaks in the absence of observations and individual risk assessment from partial observations, on both irreversible and recurrent compartmental models; finally, Section \ref{sec:conclusions} draws the conclusions and highlights future directions to be investigated.

	\section{Methodology}\label{sec:methods}
    
	\subsection{Definition of the stochastic epidemic model and observations}\label{sec:stochmodel}
    
	For simplicity, we will present the proposed method on one of the simplest epidemic models, namely the discrete-time stochastic Susceptible-Infected (SI) model. The generalization to more general epidemic models is discussed in Section~\ref{sec:generalization_othermodels}. It is thus considered the dynamics of the SI model on a population of $N$ individuals over a temporal window of $T$ time steps (which, for the sake of concreteness, will be called ``days''). The daily contacts are directly encoded in the set of parameters specifying the infection transmission, with $\lambda_{ij}^{t}$ being the infection probability along the directed edge from individual $i$ to individual $j$ at time $t$; conversely, we set $\lambda_{ij}^{t}=\lambda_{ji}^{t}=0$ if $i$ and $j$ are not in contact at time $t$.
	The epidemic state of the population at time $t$ is represented by a binary vector ${\bm x}^{t}=(x_{1}^{t},\dots,x_{N}^{t})$, with $x_{i}^{t}=S$ (resp. $x_{i}^{t}=I$) meaning that $i$
	is a Susceptible (resp. Infected) individual at time $t$. We include in the model a small self-infection probability $\varepsilon_i^t$. In general, self-infections can model scenarios where the contact network is only partially known, in such a way to avoid incompatibility between the inference model and observations arising from unknown transmission channels; however, a small self-infection can also be useful to heal pathological cases that can arise in the presence of noise-less observations (more details in Appendix \ref{app:leaves}).
	The epidemic model is assumed to be	Markovian, i.e. the time evolution of the probability $p_{t}\left({\bm x}^t\right)$ that the population is in state ${\bm x}^{t}$ at time $t$ is given in terms of the following master equation 
	\begin{equation}\label{eq:mastereq}
		p_{t+1}\left({\bm x}^{t+1}\right)=\sum_{{\bm x}^{t}\in \{S,I\}^N}W\left({\bm x}^{t+1}| {\bm x}^{t}\right) p_{t}\left({\bm x}^{t}\right),
	\end{equation}
	with transition rates factorized over nodes $W\left({\bm x}^{t+1}| {\bm x}^{t}\right)=\prod_{i}W_{i}\left(x_{i}^{t+1}| {\bm x}^{t}\right)$
	where
		\begin{align}
            W_{i}\left(x_{i}^{t+1}=S| {\bm x}^{t}\right) & =  \delta_{x_i^t,S}\alpha_i^t e^{h_i^t}, \\
			W_{i}\left(x_{i}^{t+1}=I| {\bm x}^{t}\right) & =  \delta_{x_i^t,I}+\delta_{x_i^t,S} \left[1-\alpha_i^t e^{h_i^t}\right]\nonumber\\
            & = 1-\delta_{x_i^t,S}\alpha_i^te^{h_i^t},
		\end{align}
    where $\delta_{x,y}$ denotes the Kronecker symbol.
    For notational simplicity, we introduced the probability of not being self-infected $\alpha_i^t=1-\varepsilon_i^t$ and the local field $h_{i}^{t}=\sum_{j=1}^N \nu_{ji}^{t}\delta_{x_{j}^{t},I}$, where $\nu_{ji}^{t}=\log\left(1-\lambda_{ji}^{t}\right)$, such that $e^{h_{i}^{t}}=\prod_{j=1}^N(1- \lambda_{ji}^{t}\delta_{x_{j}^{t},I})$ is the probability of not being infected by the neighbors. Then the probability of the epidemic spreading ${\bm x}= \{{\bm x}^0, {\bm x}^1,\dots, {\bm x}^T\}$ can be written, under the Markov assumption, in the following form
    \begin{equation}\label{eq:prior}
        p({\bm X})=\prod_{t=0}^T p_{t}({\bm x^{t}})=\prod_i p_0(x_i^0) \prod_{t=0}^{T-1}W_i\left(x_i^{t+1}| {\bm x}^t\right),
    \end{equation}
	where, at the initial time $t=0$, the individuals are assumed to be independent and identically distributed according to $p_0(\boldsymbol{x}^0)$.
	The likelihood of the model can be defined by a set $\boldsymbol{\mathcal{O}}=\{{\bm O}_i\}_{i=1,\dots,N}$ of site-independent observations. We assume that each node $i$ can be observed multiple times across different time steps. Observations are factorized over time as ${\bm O}_i = (O_i^0, \dots, O_i^T)$, where $O_i^t = \varnothing$ denotes the absence of observation at time $t$. A more general scenario admits uncertainty on the outcome of the tests, the latter being eventually quantified by false positive rate $\fpr$ and/or false negative rates $\fnr$. The likelihood of an observation $O_i^t$ on node $i$ at time $t$ can be written as
    \begin{equation}
        p(O_i^t| x_i^t)=\begin{cases}
				\left(1-\fpr\right)\delta_{x_{i}^t,S}+\fnr\delta_{x_{i}^t,I} & \text{if }O_{i}^{t} = S\\
				\fpr\,\delta_{x_{i}^{t},S}+\left(1-\fnr\right)\delta_{x_{i}^{t},I} & \text{if }O_{i}^{t}= I\\
                1 & \text{if }O_i^t=\varnothing
			\end{cases}\label{eq:likelihood_singleobs}
    \end{equation}
    The likelihood over the full set of observations $\bm{\mathcal{O}}$ can be written as $p\left(\bm{\mathcal{O}}| {\bm X}\right)=\prod_{i=1}^N \prod_{t=0}^T p\left(O_i^t| x_i^T\right)$.
	In the case of perfectly accurate tests, in which $\fpr=\fnr=0$, the effect of the observations is to enforce the dynamical trajectories to be compatible with the observed states.

	The posterior probability of the trajectory ${\bm X}$ can be expressed using Bayes' theorem as follows
    \begin{subequations}
    \begin{align}
        p\left({\bm X}| \boldsymbol{\mathcal{O}}\right) &=  \frac{1}{p\left(\boldsymbol{\mathcal{O}}\right)}p\left({\bm X}\right) p\left( \boldsymbol{\mathcal{O}} |  \bm X  \right) \\
        &\propto \prod_{i=1}^N p_0(x_i^0)p(O_i^T| x_i^T)\nonumber\\
        &\quad \times \prod_{t=0}^{T-1}W_i\left(x_i^{t+1}| {\bm x}^t\right)p(O_i^t| x_i^t).\label{eq:posterior2}
    \end{align}
    \end{subequations}
		
    The Bayesian inference problem consists in evaluating marginals of the posterior distribution $p\left({\bm X}| \boldsymbol{\mathcal{O}}\right)$, such as the quantity $p\left( x_i^t=x| \boldsymbol{\mathcal{O}}\right)$ representing the posterior probability that individual $i$ is in state $x\in\{S,I\}$ at time $t$ given the set of the available observations $\boldsymbol{\mathcal{O}}$. The posterior distribution is, in general, intractable but it is the starting point for the derivation of approximate inference methods.  
        
    \subsection{The Dynamic Cavity Equations for the SI model with observations}

    The posterior probability in Eq.~\eqref{eq:posterior2} can be interpreted as a graphical model for dynamical trajectories defined on the contact network. In this context, a Belief Propagation (BP) approach was proposed in Refs.~\cite{altarelli2014bayesian,altarelli2014patient}, and has since been employed for epidemic inference. This approach involves defining messages between neighboring nodes that depend on their pairwise dynamical trajectories, allowing short loops induced by the dynamical constraints in the factor graph to be disentangled.

    In this work, we propose an alternative formulation based on a cavity argument, which yields a set of equations for the marginal probability of a pair of variable-field trajectories on each node in the cavity graph. This formulation, referred to as the {\it observation-reweighted dynamic cavity (DC) equations}, provides a rigorous approach to describing the posterior distribution on sparse graphs. Starting from the posterior probability in Eq.~\eqref{eq:posterior2}, the derivation (see Appendix~\ref{app:BP_derivation} and Refs.~\cite{altarelli2014bayesian,altarelli2014patient,baker2021epidemic}) leads to the following set of equations for the cavity messages $\hat{c}_{i\setminus j}\left({\bm x}_i,{\bm s}_i|  \boldsymbol{\mathcal{O}}\right)$:
    \begin{align}
        \hat{c}_{i\setminus j}\left({\bm x}_{i},{\bm s}_{i}|  \boldsymbol{\mathcal{O}} \right) &  \propto p_0(x_{i}^{0}) p\left( {\bm O}_i |  {\bm x}_i \right) \sum_{{\bm x}_{\partial i \setminus j}} e^{\hat{S}_{i\setminus j}}  \nonumber \\
        &\quad \times \prod_{k\in\partial i\setminus j}  \hat{c}_{k\setminus i}\left({\bm x}_{k},{\bm \nu}_{ik} {\bm x}_{i}| \boldsymbol{\mathcal{O}}  \right) , \label{eq:dyncav}
    \end{align}
    where $\partial i\setminus j$ is the set of neighboring indices of $i$ except for $j$, $\sum_{{\bm x}_{\partial i \setminus j}}=\prod_{k\in\partial i\setminus j} \prod_{t=0}^{T}\sum_{x_k^t\in\{S,I\}}$, and $ \hat{S}_{i\setminus j}$ is the cavity dynamical action of the spreading dynamics under study, which for the SI model can be written as
    \begin{align}
        \hat{S}_{i\setminus j} & = \sum_{t=0}^{T-1}\log \left[ \alpha_i^t e^{s_i^t+\sum_{k\in\partial i \setminus j}\nu_{ki}^t\delta_{x_k^t,I}} \left( \delta_{x_i^{t+1},x_i^t}-\delta_{x_i^{t+1},I} \right) \nonumber \right. \\ 
        & \left. \quad+ \delta_{x_i^{t+1},I} \right]. \label{eq:cav_dyn_action}
    \end{align}
    
    For the sake of brevity, we introduced the compact representations ${\bm \nu}_{ik} {\bm x}_{i}=({\nu}_{ik}^0 \delta_{{x}_{i}^0,I},\dots, {\nu}_{ik}^T \delta_{{x}_{i}^T,I})$ and $p({\bm O}_i| {\bm x}_i)=\prod_{t=0}^Tp(O_i^t| x_i^t)$. The terms $\hat{c}_{i\setminus j}\left({\bm x}_{i},{\bm s}_{i}|  \boldsymbol{\mathcal{O}}\right)$ are the cavity messages, depending on the trajectory ${\bm x}_{i}$ of individual $i$ and the local field trajectory ${\bm s}_i$, which acts as a proxy for the trajectory of the missing neighboring node $j$ (more precisely, ${\bm s}_i \propto {\bm \nu}_{ji} {\bm x}_j$).
    
    The observation-reweighted dynamic cavity equations can be interpreted as a message-passing procedure, analogous to Belief Propagation (BP), where messages are iteratively exchanged between nodes until convergence. Importantly, as in BP, the normalization of the messages at each step is not critical and can be adjusted after convergence. Additionally, these equations recover the BP messages through a change of variables, with ${\bm s}_i$ effectively replacing the explicit dependence on the neighboring node’s trajectory. A detailed discussion of this equivalence is provided in Appendices~\ref{app:BP_derivation} and \ref{app:derivation_dynamic_cavity}.

	While the dynamic cavity equations with variable-field trajectories were originally proposed only for time-forward binary spin dynamics \cite{neri2009cavity,kanoria2011majority}, those in Eq.~\eqref{eq:dyncav} also account for probabilistic reweighting due to the observations. 
		
    Finally, completing the cavity and computing the total marginal over $i$ gives the posterior marginal probability of one-site trajectories 
    \begin{align}
        c_{i}\left({\bm x}_{i} |  \boldsymbol{\mathcal{O}} \right) & = \frac{p_0(x_{i}^{0})}{\mathcal{Z}_{i}\left(\boldsymbol{\mathcal{O}}\right)} p\left( {\bm O}_i |  {\bm x}_i \right) \sum_{{\bm x}_{\partial i}}e^{S_{i}} \nonumber \\
        &\quad \times  \prod_{k\in\partial i} \hat{c}_{k\setminus i}\left({\bm x}_{k},{\bm \nu}_{ik} {\bm x}_{i}| \boldsymbol{\mathcal{O}}  \right), 
    \end{align}
    where the summation runs over the trajectories of all neighbors of $i$, i.e, ${\bm x}_{\partial i}=\{{\bm x}_k\}_{k \in \partial i}$. The term $S_i$ represents the local dynamical action, which for the SI model is given by:
    \begin{align}
        S_i & = \sum_{t=0}^{T-1}\log\left[ \alpha_i^te^{\sum_{k\in\partial i}\nu_{ki}^t\delta_{x_k^t,I}} \left( \delta_{x_i^{t+1},x_i^t}-\delta_{x_i^{t+1},I} \right)\nonumber \right.\\
        &\left. \quad + \delta_{x_i^{t+1},I} \right].
    \end{align}
    The normalization factor is defined as
    \begin{align}
        \mathcal{Z}_{i}\left(\boldsymbol{\mathcal{O}}\right)&=\sum_{{\bm x}_i} p_0(x_{i}^{0}) p\left( {\bm O}_i |  {\bm x}_i \right) \sum_{{\bm x}_{\partial i}}e^{S_{i}} \nonumber \\
        &\quad \times  \prod_{k\in\partial i} \hat{c}_{k\setminus i}\left({\bm x}_{k},{\bm \nu}_{ik} {\bm x}_{i}| \boldsymbol{\mathcal{O}}  \right).
    \end{align}

    \subsection{Small-coupling expansion}\label{sec:smallcoupling}

    To simplify the cavity equations in Eq.~\eqref{eq:dyncav}, we express the cavity messages in terms of the conjugate field trajectories $\hat{{\bm h}}_i= (\hat{h}_i^0,\dots, \hat{h}_i^T)$ by introducing their Fourier transform with respect to ${\bm s}_i$ (see Appendix~\ref{app:derivation_dynamic_cavity}). This yields the following expression for the dynamic cavity equations:
    \begin{align}\label{eq:dyncav_fourier}
        c_{i\setminus j}({\bm x}_{i},\hat{{\bm h}}_{i} |  \boldsymbol{\mathcal{O}}) &  = \frac{p_0\left(x_{i}^{0}\right)}{\mathcal{Z}_{i\setminus j}(\boldsymbol{\mathcal{O}})} p\left( {\bm O}_i |  {\bm x}_i \right) e^{S_i^0}\nonumber \\
        &  \times \prod_{k\in\partial i\setminus j} \sum_{{\bm x}_{k}}\int D\hat{{\bm h}}_{k}\, c_{k\setminus i}({\bm x}_{k},\hat{{\bm h}}_{k} |  \boldsymbol{\mathcal{O}})e^{S_{ik}^{int}},
    \end{align}
    where $\sum_{{\bm x}_i}=\prod_{t=0}^T \sum_{x_i^t\in \{S,I\}}$ and $\int D\hat{{\bm h}}_i = \prod_{t=0}^T \int_{-\infty}^{\infty} d\hat{h}_i^t$. We used the following definition of the Fourier transform and its inverse
    \begin{align}
        c({\bm x},\hat{{\bm h}}|\boldsymbol{\mathcal{O}}) & = \int D{\bm s} \left(\prod_{t=0}^T\frac{e^{{\rm i} s^t \hat{h}^t}}{2\pi}\right)\hat{c}({\bm x},{\bm s}|\boldsymbol{\mathcal{O}}),\label{eq:fourier_transf}\\
        \hat{c}({\bm x},{\bm s}|\boldsymbol{\mathcal{O}}) & = \int D\hat{{\bm h}} \left(\prod_{t=0}^T e^{-{\rm i} s^t\hat{h}^t}\right)c({\bm x},\hat{{\bm h}}|\boldsymbol{\mathcal{O}})\label{eq:inverse_fourier_transf}.
    \end{align}

    The local non-interacting and interacting actions are defined by the epidemic model under study. For the SI model they can be written respectively as
        \begin{align}
        S_i^0 &= \sum_{t=0}^{T-1}\log\left[ \delta(\hat{h}_i^t-{\rm i})\alpha_i^t\left( \delta_{x_i^{t+1},x_i^t}-\delta_{x_i^{t+1},I} \right)\right. \nonumber\\
        &\quad\left.+\delta(\hat{h}_i^t)\delta_{x_i^{t+1},I} \right]\label{eq:local_nonint_action},\\
        S_{ik}^{int}=S_{ki}^{int}&=-{\rm i}\sum_{t=0}^{T-1}\left( \delta_{x_i^t,I}\nu_{ik}^t\hat{h}_k^t + \delta_{x_k^t,I}\nu_{ki}^t\hat{h}_i^t \right)\label{eq:local_int_action}.
    \end{align}
    Here $\delta(x)$ denotes the Dirac delta function. To ensure proper normalization of the cavity messages $c_{i\setminus j}({\bm x}_i, \hat{{\bm h}}_i | \boldsymbol{\mathcal{O}})$, we define the normalization factor $\mathcal{Z}_{i\setminus j}(\boldsymbol{\mathcal{O}})$ such that
    \begin{equation}\label{eq:norm_cond_dyncav}
        \sum_{{\bm x}_i} \int D\hat{{\bm h}}_i\, c_{i\setminus j}({\bm x}_i, \hat{{\bm h}}_i | \boldsymbol{\mathcal{O}}) = 1.
    \end{equation}
    This condition ensures that $c_{i\setminus j}({\bm x}_i, \hat{{\bm h}}_i | \boldsymbol{\mathcal{O}})$ represents a properly normalized quasi-probability distribution. Consequently, the expectation of any generic function $f({\bm x}_i, \hat{{\bm h}}_i)$ over the cavity distribution can be computed as
    \begin{equation}
        \left\langle f({\bm x}_i, \hat{{\bm h}}_i) \right\rangle_{i\setminus j}^{\boldsymbol{\mathcal{O}}} = \sum_{{\bm x}_i} \int D\hat{{\bm h}}_i\, c_{i\setminus j}({\bm x}_i, \hat{{\bm h}}_i | \boldsymbol{\mathcal{O}}) f({\bm x}_i, \hat{{\bm h}}_i).
    \end{equation}
    
    By applying the inverse Fourier transform to Eq.~\eqref{eq:norm_cond_dyncav}, we find that the normalization condition is equivalent to assuming that the neighboring node $j$ remains in the susceptible state throughout the process in the cavity graph. This corresponds to setting the local field to zero, ${\bm s}_i = \mathbf{0}$
    \begin{equation}
        \sum_{{\bm x}_i} \int D\hat{{\bm h}}_i\, c_{i\setminus j}({\bm x}_i, \hat{{\bm h}}_i | \boldsymbol{\mathcal{O}}) = \sum_{{\bm x}_i} \hat{c}_{i\setminus j}({\bm x}_i, {\bm s}_i = {\bm 0} | \boldsymbol{\mathcal{O}}) = 1.
    \end{equation}
    
    Under the small-coupling approximation, the dynamic cavity method assumes that the absence of node $j$ in the cavity graph implies no influence on the dynamics of node $i$, as $j$ is treated as always susceptible.
    In contrast, the original dynamic cavity method \cite{kanoria2011majority} explicitly incorporates the fixed trajectories of the missing neighbor $j$ and updates cavity messages through a self-consistent dynamic message-passing algorithm. However, this approach has a computational complexity that can potentially grow exponentially with the number of time steps $T$. On the one hand, for non-recurrent dynamics this complexity can be mitigated by reformulating the cavity messages in terms of flipping times \cite{lokhov2015dynamic}, offering a more efficient representation. On the other hand, this flipping-time representation is not applicable to recurrent dynamics.
	Choosing the normalization to be independent of the trajectory of individual $j$ is highly advantageous  for developing an approximation that does not explicitly rely on the epidemic trajectories of both $i$ and $j$. However, the main drawback of this choice is that some particular epidemic trajectories of individual $i$, imposed by the observations, can only be explained by the statistical model if a non-zero self-infection probability is introduced. In Appendix~\ref{app:leaves}, we provide an example of this potential issue that is simple enough to be discussed analytically and show how the presence of a self-infection probability effectively resolves it. 
    
    In the spirit of Plefka's approach \cite{Plefka_1982,Braviplefkadynamics_2016} and high-temperature expansions \cite{georges1991expand,maillard2019high}, one can perform a formal expansion of the interactive exponential term. Since the argument of the exponential is linear in the parameters $\nu_{ik}^t$ and $\nu_{ki}^t$, truncating the expansion at some finite order can be understood as a small-coupling approximation of the dynamic cavity equations, which in the case of epidemic processes corresponds to a small infectivity approximation (i.e. $\lambda_{ij}^t \ll 1$ for all $i,j$ and $t$). Truncating the Taylor series at the first order, after re-exponentiation of the interacting part we obtain an approximate expression for the cavity messages
    \begin{align}
        c_{i\setminus j}({\bm x}_{i},\hat{{\bm h}}_{i} |  \boldsymbol{\mathcal{O}}) & \approx \frac{p_0\left(x_{i}^{0}\right)}{\tilde{\mathcal{Z}}_{i\setminus j}(\boldsymbol{\mathcal{O}})} p\left( {\bm O}_i |  {\bm x}_i \right) e^{S_i^0 + \sum_{k\in\partial i \setminus j}\left\langle S_{ik}^{int} \right\rangle_{k\setminus i}^{\boldsymbol{\mathcal{O}}}}\nonumber \\
        & \coloneq \tilde{c}_{i\setminus j}({\bm x}_{i},\hat{{\bm h}}_{i} |  \boldsymbol{\mathcal{O}}),\label{eq:dyncav_expanded}
    \end{align}
	where also the normalization constant is consistently approximated to $\tilde{\mathcal{Z}}_{i\setminus j}\left(\boldsymbol{\mathcal{O}}\right)$. The averaged interacting action is a function of both the variable-conjugate field trajectories $({\bm x}_i,\hat{\bm h}_i)$ and of a set of local one-time cavity averages coming from the neighbors, namely $m_{k\setminus i}^t(\boldsymbol{\mathcal{O}})=\langle \delta_{x_k^t,I}\rangle_{k\setminus i}^{\boldsymbol{\mathcal{O}}}$ and $\mu_{k\setminus i}^t(\boldsymbol{\mathcal{O}})=\langle -{\rm i}\hat{h}_k^t\rangle_{k\setminus i}^{\boldsymbol{\mathcal{O}}}$. For the SI model, this yields
    \begin{equation}
        \left\langle S_{ik}^{int} \right\rangle_{k\setminus i}^{\boldsymbol{\mathcal{O}}}=\sum_{t=0}^{T-1}\left( \delta_{x_i^t,I}\nu_{ik}^t\mu_{k\setminus i}^t (\boldsymbol{\mathcal{O}}) -{\rm i} m_{k\setminus i}^t(\boldsymbol{\mathcal{O}})\nu_{ki}^t\hat{h}_i^t\right).\label{eq:expanded_interactive_action}
    \end{equation}
    
    The dynamic cavity equations in Eq.\eqref{eq:dyncav_expanded}, derived under the small-coupling approximation, can be closed self-consistently by evaluating the cavity averages using the approximated distribution functions
    \begin{align}
        m_{k\setminus i}^{t}\left(\boldsymbol{\mathcal{O}}\right) & \approx  \sum_{{\bm x}_{k}}\int D\hat{{\bm h}}_{k}\tilde{c}_{k\setminus i}({\bm x}_{k},\hat{{\bm h}}_{k}| \boldsymbol{\mathcal{O}})\delta_{x_{k}^{t},I}\nonumber \\
        & \approx \sum_{{\bm x}_{k}}\hat{\tilde{c}}_{k\setminus i}\left({\bm x}_{k},{\bm s}_{k}={\bm 0} | \boldsymbol{\mathcal{O}} \right)\delta_{x_{k}^{t},I}, \label{eq:m_ki_t_def}\\
        \mu_{k\setminus i}^{t}\left(\boldsymbol{\mathcal{O}}\right) & \approx  \sum_{{\bm x}_{k}}\int D\hat{{\bm h}}_{k}\tilde{c}_{k\setminus i}({\bm x}_{k},\hat{{\bm h}}_{k} |  \boldsymbol{\mathcal{O}})\left(-{\rm i}\hat{h}_{k}^{t}\right)\nonumber \\
        & \approx  \sum_{{\bm x}_{k}}\left.\frac{\partial}{\partial s_{k}^{t}}\hat{\tilde{c}}_{k\setminus i}\left({\bm x}_{k},{\bm s}_{k} | \boldsymbol{\mathcal{O}} \right)\right| _{{\bm s}_{k}={\bm 0}}, \label{eq:mu_ki_t_def}
    \end{align}
    where $\hat{\tilde{c}}_{k\setminus i}\left({\bm x}_{k},{\bm s}_{k} | \boldsymbol{\mathcal{O}} \right)$ is the approximated cavity message Eq.~\eqref{eq:dyncav}, obtained by the inverse Fourier transform of Eq.~\eqref{eq:dyncav_expanded}.
	
	The quantity $m_{k\setminus i}^{t}$ measures the average probability that node $k$ is infected at time $t$ in the absence of the interaction with $i$, while $\mu_{k\setminus i}^{t}$ is of less
	intuitive interpretation, as it  measures the mean of fluctuations around the unperturbed single-site statistics in the cavity graph. A direct calculation of the quantity $\sum_{{\bm x}_{k}}\int D\hat{{\bm h}}_{k} (-{\rm i}\hat{h}_k^t)\tilde{c}_{k\setminus i}({\bm x}_{k},\hat{{\bm h}}_{k}| \boldsymbol{\mathcal{O}}=\varnothing)$  shows that, in the absence of observations, $\mu_{k\setminus i}^{t}\left(\boldsymbol{\mathcal{O}}=\varnothing\right)=0$. This result, due to the causality of the dynamical process, does not hold anymore when some observations are included.
    
	For notational convenience, we will omit the explicit dependence of all marginals, normalization constants, and cavity averages on the observation set $\boldsymbol{\mathcal{O}}$ in the next sections. Similarly, approximated quantities will no longer be distinguished by the $\tilde{\cdot}$ notation, as all references will pertain to their approximated forms unless explicitly stated otherwise.

	\subsection{Small-Coupling Dynamic Cavity (SCDC) approximation}

    By implementing the closure assumption, where the cavity averages $m_{i\setminus j}^t$ and $\mu_{i\setminus j}^t$ are replaced by their expectations over the approximated cavity messages Eq.~\eqref{eq:dyncav_expanded}, the small-coupling expansion yields a closed set of dynamic equations for the cavity averages. This becomes evident when the approximated cavity messages Eq.~\eqref{eq:dyncav_expanded} are substituted into the expressions for the cavity averages Eqs.~\eqref{eq:m_ki_t_def}~and~\eqref{eq:mu_ki_t_def}. After integrating out the conjugate fields ${\bm h}_i$
    \begin{align}
        m_{i\setminus j}^t(\boldsymbol{\mathcal{O}}) &= \sum_{{\bm x}_i}\delta_{x_i^t,I}\frac{p_0(x_i^0)}{\mathcal{Z}_{i\setminus j}(\boldsymbol{\mathcal{O}})} p\left( {\bm O}_i |  {\bm x}_i \right) e^{S_{i\setminus j}^m}\label{eq:noncausalMFm},\\
        \mu_{i\setminus j}^t(\boldsymbol{\mathcal{O}}) &= \sum_{{\bm x}_i}\frac{p_0(x_i^0)}{\mathcal{Z}_{i\setminus j}(\boldsymbol{\mathcal{O}})} p\left( {\bm O}_i |  {\bm x}_i \right) e^{S_{i\setminus j}^\mu}\label{eq:noncausalMFmu},
    \end{align}
    where we introduced the two terms
    \begin{align}
        S_{i\setminus j}^m & = \sum_{t=0}^{T-1}\log\left[\alpha_i^te^{\sum_{k\in\partial i \setminus j}\nu_{ki}^tm_{k\setminus i}^t}\left(\delta_{x_i^{t+1},x_i^t}-\delta_{x_i^t,I}\right)\nonumber\right.\\
        &\left.\quad+\delta_{x_i^{t+1},I}\right]+\sum_{t=0}^{T-1}\sum_{k\in\partial i\setminus j}\delta_{x_i^t,I}\nu_{ik}^t\mu_{k\setminus i}^t,\\
        S_{i\setminus j}^\mu & = \sum_{t=0}^{T-1}\log\left[\alpha_i^t\left(\delta_{x_i^{t+1},x_i^t}-\delta_{x_i^t,I}\right)\right]\nonumber\\
        &\quad+\sum_{t=0}^{T-1}\sum_{k\in\partial i\setminus j}\left(\nu_{ki}^tm_{k\setminus i}^t+\delta_{x_i^t,I}\nu_{ik}^t\mu_{k\setminus i}^t\right)  
    \end{align}
    The normalization constant is chosen to ensure that the time-dependent
    quantity $m_{i\setminus j}^{t}$ represents the probability of finding individual $i$ infected at time $t$ in the cavity graph, that is 
    \begin{align}
        \mathcal{Z}_{i\setminus j}(\boldsymbol{\mathcal{O}}) & =\sum_{{\bm x}_i}p_0(x_i^0) p\left( {\bm O}_i |  {\bm x}_i \right) e^{S_{i\setminus j}^m}.\label{eq:noncausalMFzij}
    \end{align}
	In addition, the total time-dependent marginal $m_{i}^{t}$ of the posterior distribution on the full graph is given by
	\begin{equation}
        m_{i}^{t}(\boldsymbol{\mathcal{O}}) = \sum_{{\bm x}_i}\delta_{x_i^t,I}\frac{p_0(x_i^0)}{\mathcal{Z}_{i}(\boldsymbol{\mathcal{O}})} p\left( {\bm O}_i |  {\bm x}_i \right) e^{S_{i}^m} \label{eq:marginals_mi_main},
	\end{equation}
	with the quantity $S_i^m$ and the normalization factor $\mathcal{Z}_i$ defined respectively as
    \begin{align}
    	S_{i}^m & = \sum_{t=0}^{T-1}\log\left[\alpha_i^te^{\sum_{k\in\partial i}\nu_{ki}^tm_{k\setminus i}^t}\left(\delta_{x_i^{t+1},x_i^t}-\delta_{x_i^t,I}\right)\nonumber\right.\\
        &\left.\quad+\delta_{x_i^{t+1},I}\right]+\sum_{t=0}^{T-1}\sum_{k\in\partial i}\delta_{x_i^t,I}\nu_{ik}^t\mu_{k\setminus i}^t,\\
        Z_i(\boldsymbol{\mathcal{O}}) & =\sum_{{\bm x}_i}p_0(x_i^0) p\left( {\bm O}_i |  {\bm x}_i \right) e^{S_{i}^m}.
    \end{align}
	
	Equations~\eqref{eq:noncausalMFm}~and~\eqref{eq:noncausalMFmu} represent a set of self-consistent equations defining a non-causal dynamic mean-field approximation that, in what follows, we will refer to as the Small-Coupling Dynamic Cavity (SCDC) method. The dynamical equations are of mean-field type since correlations are neglected, but in the presence of observations, they describe a non-causal dynamical process. Because of the cavity construction, the fundamental unknown of the equations, the one-time cavity marginals $m_{i\setminus j}^t$ and the one-time cavity fields $\mu_{i\setminus j}^t$, are defined by means of local self-consistent conditions, the solution of which can be sought through an iterative message-passing scheme. A computational bottleneck of Eqs.~\eqref{eq:noncausalMFm}~and~\eqref{eq:noncausalMFmu} is represented by the partial trace over single-site trajectories ${\bm x}_i$, that requires $O(2^T)$ operations, meaning that a complete update of all cavity quantities requires $O(2| E| T 2^T)$, where $| E| $ is the total number of non-zero weighted directed edges on the interaction graph. An efficient algorithmic implementation of the SCDC equations is proposed in the next Section. It exploits a transfer-matrix approach to perform the trace over the trajectory ${\bm x}_i$ keeping fixed all quantities $\{m_{k\setminus i}^t\}$ and $\{\mu_{k\setminus i}^t\}$ for all $t=0,\dots,T$ and $k\in \partial i\setminus j$, which play the role of the parameters in a ``temporal'' one-dimensional discrete probabilistic model defined on node $i$.
	
	\section{Efficient formulation of the SCDC equations\label{sec:efficientformulation}}
	
	The starting point of the derivation is the cavity normalization constant in Eqs.~\eqref{eq:noncausalMFm}~and~\eqref{eq:noncausalMFmu}, that can be written starting from Eq.~\eqref{eq:noncausalMFzij} as
	\begin{align}
		\mathcal{Z}_{i\setminus j} & =  \sum_{{\bm x}_{i}}p_0(x_{i}^{0}) \prod_{t=0}^{T-1}M_{x_{i}^{t}x_{i}^{t+1}}^{i\setminus j}  p\left(O_{i}^{T}|  x_{i}^{T}\right),\label{eq:partition_function}
	\end{align}
	where the  ``transfer matrix'' $M_{x^{t}x^{t+1}}^{i\setminus j}$ is defined as follows
	\begin{widetext}
		\begin{equation}
			M_{x_i^{t}x_i^{t+1}}^{i\setminus j}  =\left(\begin{array}{cc}
				M_{t,SS}^{i\setminus j} & M_{t,SI}^{i\setminus j}\\
				M_{t,IS}^{i\setminus j} & M_{t,II}^{i\setminus j}
			\end{array}\right) =  \left(\begin{array}{ccc}
				\alpha_i^te^{\sum_{k\in\partial i\setminus j}m_{k\setminus i}^{t}{\nu}_{ki}^{t}}p\left(O_{i}^{t}|  S\right) & &\left(1-\alpha_i^te^{\sum_{k\in\partial i\setminus j}m_{k\setminus i}^{t}{\nu}_{ki}^{t}}\right) p\left(O_{i}^{t}|  S\right)\\
				0 & & e^{\sum_{k\in\partial i\setminus j}\nu_{ik}^{t}\mu_{k\setminus i}^{t}}  p\left(O_{i}^{t}|  I\right)
			\end{array}\right).
			\label{eq:transfermatrix}
		\end{equation}
    \end{widetext}
	
	The approximate probability that an individual $i$ is infected at time $t$ in the cavity graph is
	\begin{equation}
			m_{i\setminus j}^{t}  =  \frac{\rho_{\rightarrow t}^{i\setminus j}\left(I\right)\rho_{t\leftarrow}^{i\setminus j}\left(I\right)}{\sum_{x_i^t\in\{S,I\}}\rho_{\rightarrow t}^{i\setminus j}\left(x_i^t\right)\rho_{t\leftarrow}^{i\setminus j}\left(x_i^t\right)}.\label{eq:mij_rho}
	\end{equation}
    The term $\rho_{\rightarrow t}^{i\setminus j}\left(x_i^t\right)$ is a cavity forward "temporal" message, obtained recursively for $t=1,\dots,T$ as
    \begin{equation}\label{eq:rhomessages_fwd}
        \rho_{\rightarrow t}^{i\setminus j}\left(x_{i}^{t}\right) =  \sum_{x_{i}^{t-1}\in \{S,I\}}\rho_{\rightarrow t-1}^{i\setminus j}\left(x_{i}^{t-1}\right)M_{x_{i}^{t-1}x_{i}^{t}}^{i\setminus j},
    \end{equation}
    with initial condition $\rho_{\rightarrow 0}^{i\setminus j}\left(x_{i}^{0}\right)  =  p\left(x_{i}^{0}\right)$, and $\rho_{t \leftarrow}^{i\setminus j}\left(x_i^t\right)$ is a cavity backward "temporal" message, that satisfy for $t=0,\dots,T-1$ the recursive equation
    \begin{equation}\label{eq:rhomessages_bwd}
            \rho_{t\leftarrow}^{i\setminus j}\left(x_{i}^{t}\right)  =  \sum_{x_{i}^{t+1}\in \{S,I\}}\rho_{t+1\leftarrow}^{i\setminus j}\left(x_{i}^{t+1}\right)M_{x_{i}^{t}x_{i}^{t+1}}^{i\setminus j}, 
    \end{equation}
    with terminal condition $\rho_{T\leftarrow}^{i\setminus j}\left(x_{i}^{T}\right) =  p\left(O_{i}^{T}|  x_{i}^{T}\right)$.
    The approximate one-time cavity field $\mu_{i\setminus j}^t$ can be computed as follows:
    \begin{equation}
            \mu_{i\setminus j}^{t} = \frac{\rho_{\rightarrow t}^{i\setminus j}\left(S\right)M_{t,SS}^{i\setminus j}\left(\rho_{t+1\leftarrow}^{i\setminus j}\left(S\right) - \rho_{t+1\leftarrow}^{i\setminus j}\left(I\right)\right)}{\sum_{x_{i}^{t},} \rho_{\rightarrow t}^{i\setminus j}\left(x_{i}^{t}\right)\rho_{t\leftarrow}^{i\setminus j}\left(x_{i}^{t}\right)}.\label{eq:muij_rho}
    \end{equation}

	The number of operations 
	necessary for the update of a single-time cavity message is now of $O(1)$ when the single-site ``temporal'' messages  $\rho_{\rightarrow s}^{i\setminus j}\left(x_i^s\right)$ and $\rho_{s \leftarrow}^{i\setminus j}\left(x_i^s\right)$ for $s=t,t+1$ are available. The latter quantities are computed by means of time-forward and time-backward update rules from the current set of cavity ``temporal'' messages and require $O(4T)$ operations. In summary, a complete update of the cavity marginals $m_{i\setminus j}^t$ and cavity fields $\mu_{i\setminus j}^t$, for every directed edge and time step, requires $O(4| E| T)$.\\ 
	A Julia-based reference implementation of the above algorithm is available at \cite{repoSCDC}.
	Notice that from Eq.~\eqref{eq:muij_rho}, $\mu_{i\setminus j}^{t} $ is zero if the time-backward cavity messages at time $t+1$ are uniform. It is shown in Section \ref{sec:timeforward} and in Appendix~\ref{app:derivationtimeforward} that this condition is satisfied when no observations are present at later times and it leads to a pure time-forward reduction of the SCDC equations. 
	Furthermore, because of the non-recurrent property of the SI model, it is possible to derive an alternative efficient formulation of the SCDC equations exploiting the infection-time representation: this is explained in detail in Appendix~\ref{app:efficientimplementation}.

    \section{Generalization to other epidemic models\label{sec:generalization_othermodels}}
	
	The method generalizes directly to models with a higher number of individual states and transitions. In particular, for the Susceptible-Infected-Susceptible (SIS) and Susceptible-Infected-Removed (SIR) models, there is only one additional transition where an infected individual $i$ can recover at time $t$ with probability $r_i^t$, with the result that the individual $i$ is either susceptible again (for the SIS model) or in a state of acquired immunity, i.e. $x_i^t=R$ (for the SIR model). A further generalization can be made for the Susceptible-Infected-Removed-Susceptible (SIRS), where each recovered individual $i$ can return to the susceptible state at time $t$ due to loss of immunity with probability $\sigma_i^t$.
	
	The only difference with the method described in the previous sections is in the expression of the transfer matrix. For the SIS model it is now given by
	\begin{widetext}	
    \begin{equation}
        M_{x_i^{t}x_i^{t+1}}^{i\setminus j} = \left(\begin{array}{cc}
            \alpha_i^te^{\sum_{k\in\partial i\setminus j}m_{k\setminus i}^{t}{\nu}_{ki}^{t}}p\left(O_{i}^{t}|  S\right)  & \left[1-\alpha_i^te^{\sum_{k\in\partial i\setminus j}m_{k\setminus i}^{t}{\nu}_{ki}^{t}}\right] p\left(O_{i}^{t}|  S\right)\\
            r_i^t e^{\sum_{k\in\partial i\setminus j}\mu_{k\setminus i}^{t}{\nu}_{ik}^{t}}p\left(O_{i}^{t}|  I\right)  & \left(1-r_i^t\right) e^{\sum_{k\in\partial i\setminus j}\nu_{ik}^{t}\mu_{k\setminus i}^{t}}  p\left(O_{i}^{t}|  I\right)
        \end{array}\right).
    \end{equation}
    In the case of the SIR model, each individual $i$ can be in three possible states $x_i^t\in \{S,I,R\}$, but the derivation done for two-state models can be repeated almost straightforwardly (see Appendix~\ref{app:SIRmodel} for details). The cavity averages are defined analogously to the SI model through Eqs.\eqref{eq:mij_rho}~and~\eqref{eq:muij_rho} where the quantities $\rho_{\rightarrow t}^{i\setminus j}\left(x_i^t\right)$ and $\rho_{t \leftarrow}^{i\setminus j}\left(x_i^t\right)$ satisfy  the set of equations Eqs.~\eqref{eq:rhomessages_fwd}~and~\eqref{eq:rhomessages_bwd}, with a $3\times 3$ transfer matrix matrix $M_{x_i^{t}x_i^{t+1}}^{i\setminus j}$ given by
    \begin{align}\label{eq:transferMatrixSIR}
        M_{x_i^{t}x_i^{t+1}}^{i\setminus j} &= \left(\begin{array}{ccc}
            \alpha_i^te^{\sum_{k\in\partial i\setminus j}m_{k\setminus i}^{t}{\nu}_{ki}^{t}}p\left(O_{i}^{t}|  S\right) & \left[1-\alpha_i^te^{\sum_{k\in\partial i\setminus j}m_{k\setminus i}^{t}{\nu}_{ki}^{t}}\right] p\left(O_{i}^{t}|  S\right) & 0\\
            0 & (1-r_i^t)e^{\sum_{k \in \partial i \setminus j} \nu_{ik}^t\mu_{k\setminus i}^t}p(\mathcal{O}_i^t| I) & r_i^te^{\sum_{k \in \partial i \setminus j} \nu_{ik}^t\mu_{k\setminus i}^t}p(\mathcal{O}_i^t| I)\\
            0 & 0 & p(\mathcal{O}_i^t| R)\end{array}\right).
    \end{align}
    The SIRS model differs from the SIR model only for the $3\times 3$ transfer matrix, which is given by
    \begin{align}
        M_{x_i^{t}x_i^{t+1}}^{i\setminus j} &= \left(\begin{array}{ccc}
            \alpha_i^te^{\sum_{k\in\partial i\setminus j}m_{k\setminus i}^{t}{\nu}_{ki}^{t}}p\left(O_{i}^{t}|  S\right) & \left[1-\alpha_i^te^{\sum_{k\in\partial i\setminus j}m_{k\setminus i}^{t}{\nu}_{ki}^{t}}\right] p\left(O_{i}^{t}|  S\right) & 0\\
            0 & (1-r_i^t)e^{\sum_{k \in \partial i \setminus j} \nu_{ik}^t\mu_{k\setminus i}^t}p(\mathcal{O}_i^t| I) & r_i^te^{\sum_{k \in \partial i \setminus j} \nu_{ik}^t\mu_{k\setminus i}^t}p(\mathcal{O}_i^t| I)\\
            \sigma_i^tp(\mathcal{O}_i^t| R) & 0 & \left( 1 - \sigma_i^t\right)p(\mathcal{O}_i^t| R)\end{array}\right).
    \end{align}
	\end{widetext}
	As long as further compartments are included with transitions being parametrized by individual-based rates, generalization of the above construction follows straightforwardly (e.g. SEIR and SEIRS models).
	
    \section{Results}\label{sec:results}
	
    In this Section, we provide numerical results to highlight the operation and capabilities of this method. We first analyze the quality of the approximation for time-forward dynamics obtained in the absence of observations; then we effectively demonstrate the role of the cavity fields $\mu^t_{i\setminus j}$ in the presence of observations. The performances of the SCDC method in various instances of epidemic inference for the SI model are presented, both on synthetic and real-world contact networks. Finally, we evaluate the inference performance of the SCDC algorithm on recurrent epidemic models, such as the SIRS, on synthetic contact networks.
    
	\subsection{Time-forward dynamics}\label{sec:timeforward}
    
	\begin{figure}
		\includegraphics{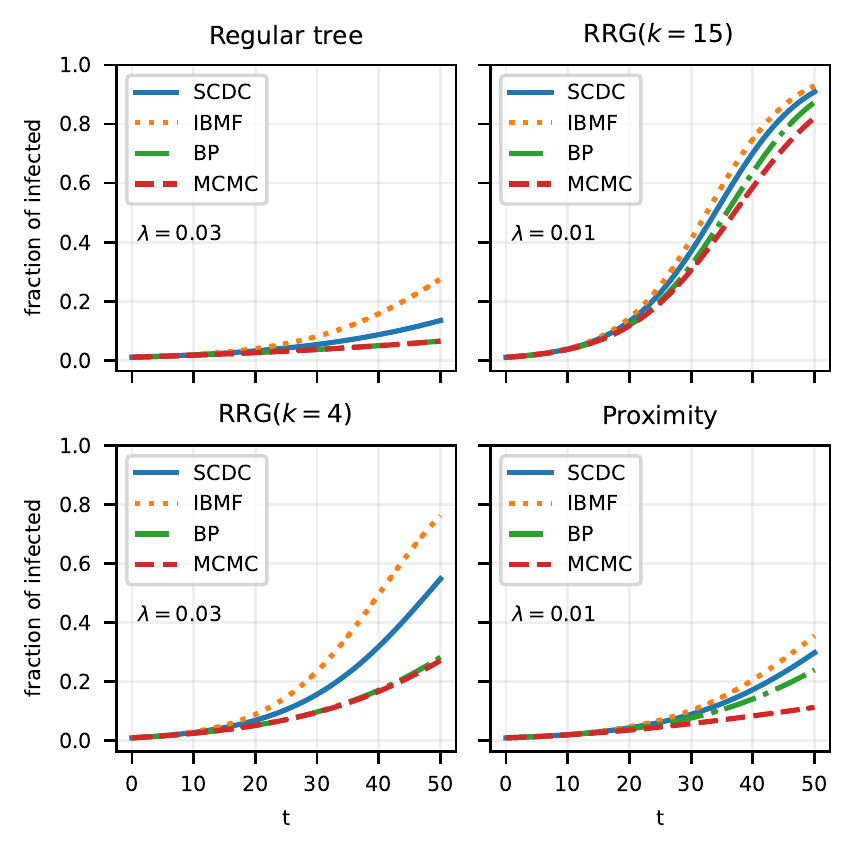}
		\caption{\textbf{Forward dynamics with SCDC and other mean field methods}. In each panel, the fraction of infected individuals is shown against the running time of the epidemic, with four different static contact graphs. Comparison is shown between SCDC, IBMF, BP and Monte Carlo simulations (with $10^4$ samples). All the links have the same infection probability $\lambda$, whose value is reported inside each panel. From left to right and top to bottom: regular tree with degree $K=4$ and $N=485$, RRG with $N=500$ and degree $K=4$, RRG with $N=500$ and $K=15$, proximity graph with $N=500$. In all cases, the probability of each individual being infected at time $t=0$ is set to $\gamma^0=5\slash N$, and the self-infection $\varepsilon_i^t$ is set to $0$.} \label{figure:res_fwd}
	\end{figure}
	
	The emerging causality-breaking of the SCDC equations is a consequence of the existence of observations at later times, that have to be taken into account in the mathematical model by a flux of information flowing backward in time and conditioning the whole history of the process. This property reflects in the existence of non-trivial values for the one-time cavity fields $\mu_{i\setminus j}^t$. 
	On the other hand, when no observation is present it is possible to show that all the cavity fields  $\mu_{i\setminus j}^t$ vanish   and, consequently, one can recover the usual causal time-forward mean-field dynamics.
	To prove this, it is convenient to start from a particular form of the update equations for the cavity marginals $m_{i\setminus j}^t$ (see Appendix~\ref{app:derivationtimeforward} for a derivation), 
    \begin{align}
			m_{i\setminus j}^{t} &  =  m_{i\setminus j}^{t-1}+\left(1-m_{i\setminus j}^{t-1}\right)  \nonumber\\
            &\quad \times \frac{\left(1-g_{i\setminus j}^{t-1}\right) \rho_{t\leftarrow}^{i\setminus j}\left(I\right)}{g_{i\setminus j}^{t-1}\rho_{t\leftarrow}^{i\setminus j}\left(S\right)+\left(1-g_{i\setminus j}^{t-1}\right) \rho_{t\leftarrow}^{i\setminus j}\left(I\right)}, \label{eq:not_forward}
	\end{align}
	where, for brevity of notation, we have defined the term $g_{i\setminus j}^t=\alpha_i^t\exp{\left(\sum_{k\in\partial i\setminus j}m_{k\setminus i}^t {\nu}_{ki}^t \right)}$. The messages $\rho_{t\leftarrow}^{i\setminus j}\left(x_i^t\right)$ represent the (non-normalized) backward probability that node $i$ has state $x_i^t \in\{S,I\}$ at time $t$, given the dynamic constraints and the observations in the future. In the absence of observations (on all nodes at all times $t'\geq t$) the backward probability is uniform, i.e. $\rho_{t\leftarrow}^{i\setminus j}\left(S\right)=\rho_{t\leftarrow}^{i\setminus j}\left(I\right)$, and Eq.~\eqref{eq:not_forward} reduces to time-forward mean-field equations, 
	\begin{align}\label{eq:mij_forward}
		m_{i\setminus j}^{t} & =  m_{i\setminus j}^{t-1}  +\left(1-m_{i\setminus j}^{t-1}\right)\nonumber \\
		& \quad \times\left[1 - \alpha_i^{t-1}e^{\sum_{k\in\partial i\setminus j}m_{k\setminus i}^{t-1}\nu_{ki}^{t-1}} \right],
	\end{align}
	and for the total marginals $m_{i}^{t}$
	\begin{align}
		\label{eq:mi_forward}
		m_{i}^{t} & =  m_{i}^{t-1}+\left(1-m_{i}^{t-1}\right)\nonumber \\
		& \quad \times\left[ 1-\alpha_i^{t-1}e^{\sum_{j\in\partial i}m_{j\setminus i}^{t-1}{\nu}_{ji}^{t-1}}\right].
	\end{align}
	It is possible to verify numerically that, in the absence of observations,  the SCDC algorithm in Eqs.~\eqref{eq:noncausalMFm}~and~\eqref{eq:noncausalMFmu} always converges to the same result obtained by running the time-forward Eqs.~\eqref{eq:mij_forward}.
	
	An intuitive form for the discrete-time IBMF dynamics, obtained by assuming independence of individual marginal probabilities of being infected, is given by the equations \cite{kiss2017mathematics,pastor2015epidemic}, 
	\begin{align}
		\label{eq:IBMF_forward}
		m_{i}^{t}  & = m_{i}^{t-1}+\left(1-m_{i}^{t-1}\right)\nonumber\\
		& \qquad \times\left[ 1- \alpha_i^t \prod_{j\in\partial i} \left(1-{\lambda}_{ji}^{t-1} m_{j}^{t-1}\right)\right].
	\end{align}
    A natural extension of Eq.~\eqref{eq:IBMF_forward} to account for cavity dynamics can be expressed as:  
    \begin{align}
        \label{eq:IBMF_forward_cav}
        m_{i}^{t}  & = m_{i}^{t-1} + \left(1 - m_{i}^{t-1}\right) \nonumber \\
        & \qquad \times \left[ 1 - \alpha_i^t \prod_{j \in \partial i} \left(1 - \lambda_{ji}^{t-1} m_{j \setminus i}^{t-1} \right) \right],
    \end{align}  
    where \(m_{j \setminus i}^t\) represents the cavity marginal probability of node \(j\) being infected, excluding the influence of node \(i\). This cavity probability evolves over time according to the following equation:  
    \begin{align}
        \label{eq:IBMF_forward_cav_recursive}
        m_{i \setminus j}^{t}  & = m_{i \setminus j}^{t-1} + \left(1 - m_{i \setminus j}^{t-1}\right) \nonumber \\
        & \qquad \times \left[ 1 - \alpha_i^t \prod_{k \in \partial i \setminus j} \left(1 - \lambda_{ki}^{t-1} m_{k \setminus i}^{t-1} \right) \right].
    \end{align}  
    This cavity formulation provides a more refined perspective by iteratively excluding the feedback loops caused by direct interactions between neighboring nodes, improving the approximation of marginal probabilities in systems with significant correlations.
	For a densely-connected graph (for which $m_{j\setminus i}^{t} \approx m_{j}^{t}$), Eqs.~\eqref{eq:IBMF_forward_cav} and \eqref{eq:IBMF_forward} reduce to the same expression.  

    Figure~\ref{figure:res_fwd} illustrates the quality of the approximation obtained using Eqs.~\eqref{eq:mij_forward}-\eqref{eq:mi_forward} for studying purely time-forward SI dynamics, in the absence of observations. The simulations were performed on well-known classes of static graphs, including regular trees (i.e., trees with degree \(K\)), Random Regular Graphs (RRGs, i.e., graphs in which every node has exactly degree \(K\)), and proximity random graphs (details provided in Sec.~\ref{sec:EPI_inference}). Comparisons are shown between the Small-Coupling Dynamic Cavity method (SCDC), the Belief Propagation (BP) algorithm, and individual-based mean-field equations (IBMF, corresponding to Eqs.~\eqref{eq:IBMF_forward}). As a reference, the results obtained with numerical sampling from \(10^4\) realizations of the exact time-forward Monte Carlo dynamics of the SI model are also reported. All methods based on mean-field approximations tend to overestimate the number of infected individuals in cases where the assumed factorization of probabilities is not exact. The BP algorithm is exact on trees (Bethe Lattice) and very accurate on sparse random graphs, where both SCDC and IBMF instead considerably overestimate the number of infected individuals. The performances of all methods are good on dense random graphs and much worse on graphs with spatial structure, such as proximity random graphs. In all cases under study, the SCDC approximation consistently gives better results than those obtained using IBMF.  

    Figure~\ref{fig:varlambda_fwd} shows how the value of \(\lambda\) influences the performance of these methods. The same types of plots as in Fig.~\ref{figure:res_fwd} are repeated for six different values of \(\lambda\), ranging from 0.05 to 0.5. In this analysis, we also included the improved cavity version of IBMF, given by Eq.~\eqref{eq:IBMF_forward_cav}, which we refer to as SCDCa. The results show that for small values of \(\lambda\), SCDC performs better than IBMF in estimating the average fraction of infected individuals on all kind of graphs. Conversely, for large values of \(\lambda\), IBMF performs slightly better than SCDC on random regular graphs (RRGs) and proximity graphs. However, the approximated version of SCDC, i.e., SCDCa, consistently outperforms both SCDC and IBMF across all cases. Despite its improved accuracy, we chose not to use this approximation for inference tasks, as it lacks sufficient theoretical justification.	
	
	\subsection{Effects of backward messages}
    
	\begin{figure*}
		\centering
		\includegraphics[width=.7\textwidth]{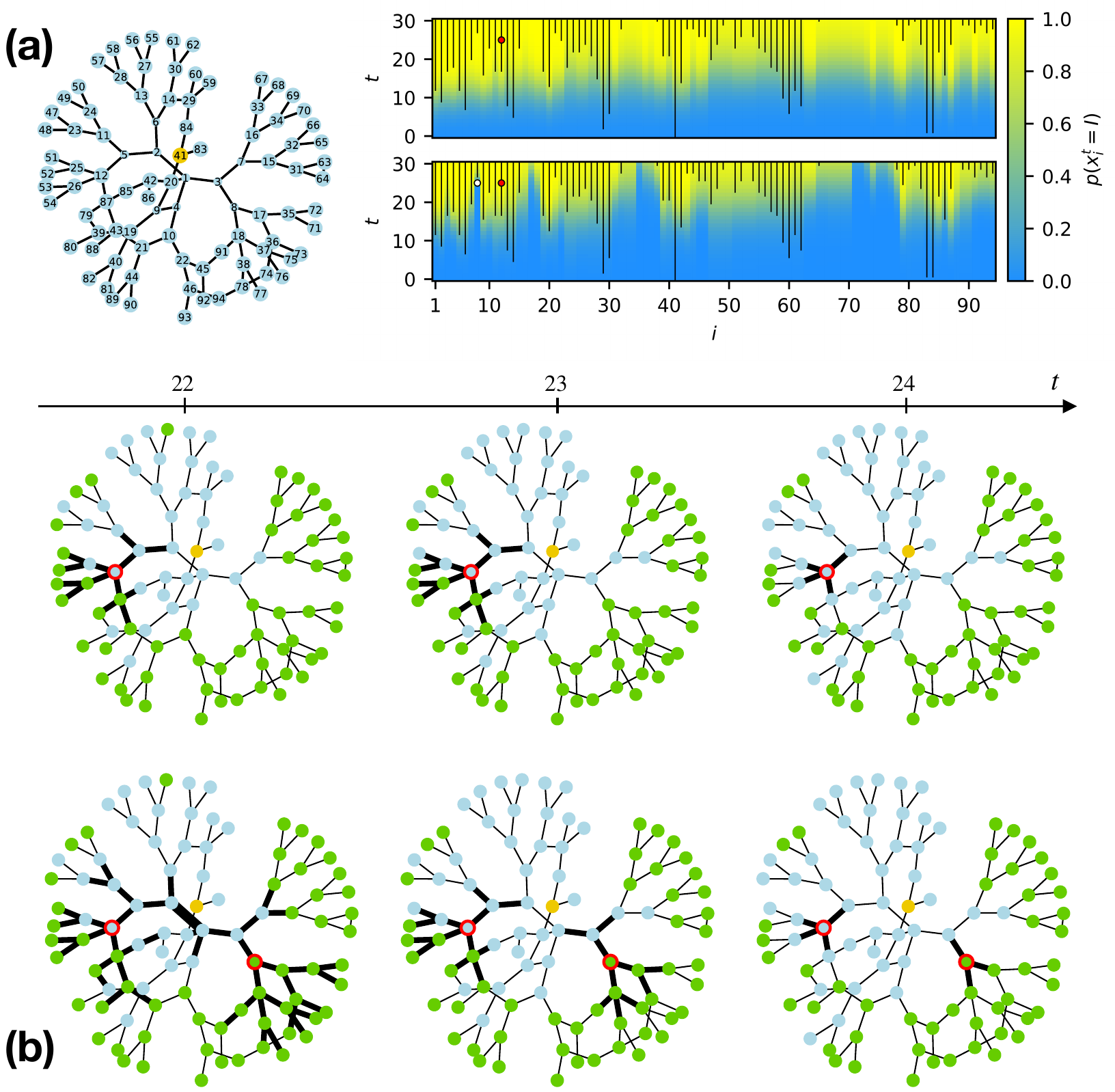}
		\caption{\textbf{Behavior of cavity fields $\mu_{i\setminus j}^t$ in the presence of observations.} Panel (a)-left: contact graph obtained randomly adding links to a tree; the epidemic source is shown as a yellow node. Panel (a)-right: inference performance of SCDC with one observation (top) and two observations (bottom). The two plots show the posterior probability of being infected as a function of time (varying on the $y$-axis) for all nodes of the contact graph (a, left) in a single epidemic outbreak generated according to a uniform SI model with $\lambda = 0.19$. The same parameter was used to perform the inference. Specifically, the top plot corresponds to one observation on node 12 at time 25, while the bottom one corresponds to two observations on nodes 12 and 8 at time 25. The corresponding state of the observations is represented by colored dots (white = $S$, red = $I$). Black vertical lines mark the true infection periods. (b) Backpropagation through time of the $\mu$ cavity fields due to observations. The observed individuals are marked by a red circle. Infected individuals are marked by a blue dot, while susceptible individuals are green. The first line of plots corresponds to one observation, while the second line corresponds to two observations. Time flows from left to right, as shown in the time arrow on the top. Thin edges correspond to vanishing $\mu$ fields, thick edges corresponds to one of the two cavity fields $\mu_{i\setminus j}^t$ or $\mu_{i\setminus j}^t$ being non-vanishing, and the thickest edges correspond to both of them being non-vanishing. Observations lead to the activation of $\mu$ fields, which propagate back in time away from the observed nodes. The portions of the graph where the $\mu$ messages are non-vanishing have better predictions, meaning that the $\mu$ fields are fundamental for Bayesian inference.}
		\label{fig:norm_mu}
	\end{figure*}
	\begin{figure*}
		\centering
		\includegraphics[width=.8\textwidth]{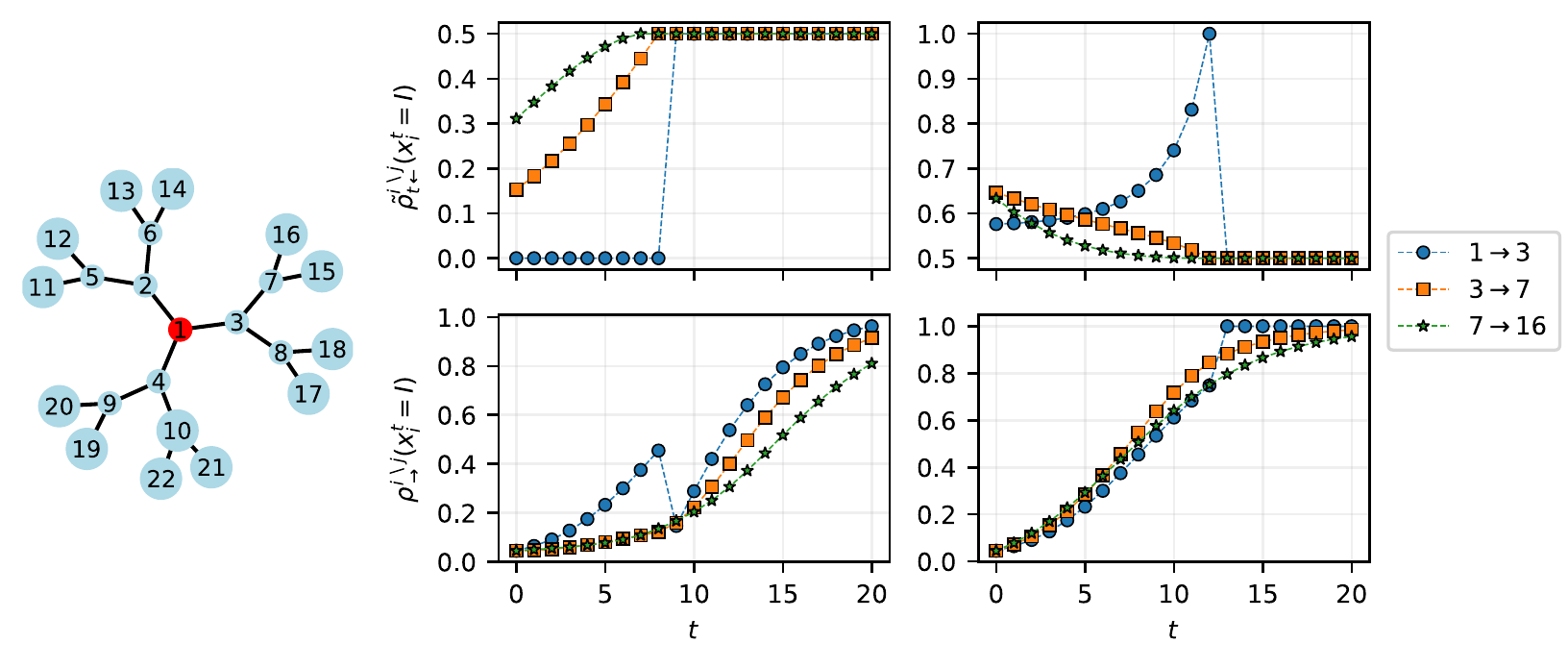}
		\caption{\textbf{Time propagation of information generated by observations.} Normalized backward (top column) and forward (bottom column) cavity messages $\tilde{\rho}_{t\leftarrow}^{i\setminus j}\left(x_i^t=I\right)$ and $\tilde{\rho}_{\rightarrow}^{i\setminus j}\left(x_i^t=I\right)$ plotted over time for different edges of a small tree. The epidemic outbreak is sampled using a uniform SI model with $\lambda=0.1$. The plot on the left shows the graph. The observed individual is marked red, while the unobserved are blue. The observed node becomes infected at time $t=10$. The plots on the left column show the messages when observing the central node being susceptible at time $t_{o}=8$. The plots on the right column show the messages when observing the central node being infected at time $t_{o}=12$.}
		\label{fig:rho_back}
	\end{figure*}
	When at least one observation is included as evidence, the $\mu$ cavity fields are non-zero and cause a time-backward propagation of information which changes the probabilistic weight of the epidemic trajectories.  In order to better understand how the method behaves in the presence of observations, we checked on which edges the absolute value of the $\mu$ cavity messages is non-vanishing when one or more observations are considered. In particular, we sampled a single epidemic outbreak from a uniform SI model with $\lambda=0.19$, on a contact graph built by randomly adding some edges between nodes of a tree. Figure \ref{fig:norm_mu}-(b) shows how the fields $\mu_{i\setminus j}^t$ propagate into the contact graph up to three times before the observation. For each edge $(i,j)$ a thick line is plotted if one of the two messages $\mu_{i\setminus j}^t$ or $\mu_{j\setminus i}^t$ is non-vanishing. A thicker line is plotted if both of them are non-vanishing. The plots are shown both for a single observation (top) and two observations (bottom). The plot on the right of Figure \ref{fig:norm_mu}-(a) shows the inference accuracy of the method for the two cases. We can see that adding an observation greatly increases the inference performance. In particular, the prediction is improved mostly on the branches of the contact tree where the new observation produces the propagation of the cavity fields $\mu$.  
	It is clear that observations lead to the activation of the $\mu$ fields, which then propagate back in time away from the observed nodes. 
	
	A more intuitive probabilistic interpretation of the role of $\mu$ cavity fields in propagating the information obtained from observations can be obtained by monitoring the temporal behavior of the (normalized) time-forward and time-backward messages $\tilde{\rho}_{\rightarrow t}^{i \setminus j}(x_i^t)=\rho_{\rightarrow t}^{i \setminus j}(x_i^t)/Z_{\rightarrow t}^{i \setminus j}$ and $\tilde{\rho}_{t\leftarrow}^{i \setminus j}=\rho_{t\leftarrow}^{i \setminus j}(x_i^t)/Z_{t\leftarrow}^{i \setminus j}$, where the normalization factor are respectively defined as
    \begin{align}
        Z_{\rightarrow t}^{i \setminus j}&=\sum_{x_{i}^{t-1}, x_{i}^{t}\in \{S,I\}}\rho_{\rightarrow t-1}^{i\setminus j}\left(x_{i}^{t-1}\right)M_{x_{i}^{t-1}x_{i}^{t}}^{i\setminus j},\\
        Z_{t\leftarrow}^{i \setminus j} & = \sum_{x_i^t, x_{i}^{t+1}\in \{S,I\}}\rho_{t+1\leftarrow}^{i\setminus j}\left(x_{i}^{t+1}\right)M_{x_{i}^{t}x_{i}^{t+1}}^{i\setminus j}.
    \end{align}
    
    Consider a realization of the SI model taking place on a small tree, as displayed in Figure \ref{fig:rho_back} (left), in which the root node gets infected at time $t=10$. An observation of the state of the root node at a time $t_o$ introduces a source of information that affects the temporal behavior of the messages $\rho_{\rightarrow t}^{i \setminus j}$ and $\rho_{t\leftarrow}^{i \setminus j}$ for all directed edges $(i,j)$ at all times. In particular, Figure \ref{fig:rho_back} (top row) shows the normalized time-backward messages $\tilde{\rho}_{t\leftarrow}^{i \setminus j}$ as function of time on a set of edges for $t_{o}=8$ (center) and $t_{o}=12$ (right). The observation of a susceptible node (center) implies that at all times before the observation, the message emerging from that node is exactly zero. Moving away from the observed node the time-backward probability is non-zero (and monotonically increasing with the spatial distance) but  monotonically decreasing with time distance from the observation. If instead the root node is observed in the infected state, the time-backward message shows an instantaneous jump to 1 at the time of observation and gradually decreases at earlier times, as the time-backward probability of being infected for the root node decreases. Moving away from the root, the messages increase as time proceeds backward indicating that surrounding nodes might have caused the infection of the observed one. Clearly, the (normalized) time-backward messages are exactly equal to 0.5 when no information is available, i.e. at later time steps compared to the observation. The normalized time-forward messages $\rho_{\rightarrow t}^{i \setminus j}$, which are shown in Figure \ref{fig:rho_back} (bottom row) exhibit a similar, though more intuitive, phenomenology. They increase from 0 to 1 with increasing time, in order to explain the epidemic cascade generated by the infectious individuals at time $t=0$. The message exiting from the central node collapses to a lower probability as soon as the individual is observed susceptible, after which it keeps increasing due to the forward spread of the dynamics. When the central node is observed infected though, the forward message exiting from it jumps to 1 only after some time-steps.
	
	\subsection{Inference performance for SI model}\label{sec:EPI_inference}
    
    \begin{figure*}
		\begin{center}
			\includegraphics[width=.8\textwidth]{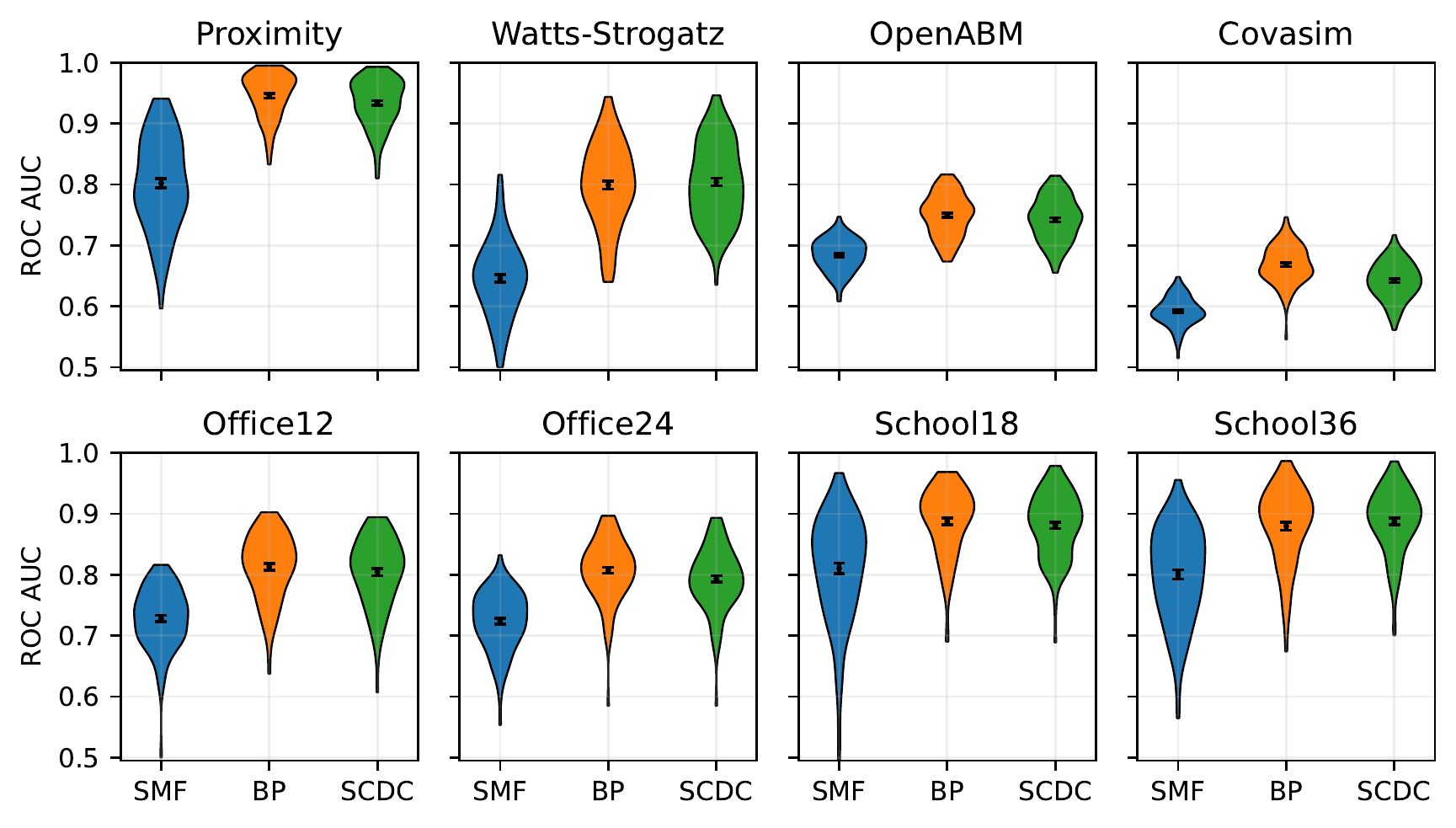}
			\caption{\textbf{Epidemic risk assessment of inference methods on SI epidemics.} Epidemic risk assessment on different classes of random contact networks. Random contact networks are shown in the top row; from left to right: soft random geometric graphs (Proximity), Watts-Strogatz random networks (Watts-Strogatz), OpenABM contact network (OpenABM) and Covasim contact network (Covasim). Real-world contact networks are shown in the bottom row. Different sizes $t_w$ of time windows were used to aggregate contact data. From left to right: office network from the InVS15 dataset with $T=12$, $t_w=24h$ (Office12) and with $T=24$, $t_w=12h$ (Office24), school network from the Thiers13 dataset with $T=18$, $t_w=6h$ (School18) and with $T=36$, $t_w=3h$ (School36). The performances of SMF, BP and SCDC are compared in the problem of classifying the infection state of $N-n_{obs}$ unobserved individuals at the last time $T$ of SI epidemic processes when a small number $n_{obs}$ of random observations are provided at the same time $T$ of the epidemic process. The performance is quantified by computing the area under the curve (AUC) for the Receiver Operating Characteristic (ROC) curves for 100 epidemic realizations for each class of contact network. The violin plots illustrate the distribution of performances. Each plot is overlaid with a black marker indicating the mean value and the associated error of the AUC across epidemic realizations. The width of the violins represents the relative frequency of data points at each value. Other parameters: soft random geometric graphs ($N=600$, $l_{\textrm{max}}=\sqrt{{2.8}/{N}}$, $n_{obs}=20$, $T=28$, $\lambda=0.08$), Watts-Strogatz graphs ($N=600$, average degree $z=4$, $p_{rw} = 0.12$, $n_{obs}=20$, $T=25$,  $\lambda=0.16$), OpenABM ($N=2000$, $n_{obs}=60$, $T=21$, $\gamma=0.026$), Covasim ($N=1000$, $n_{obs}=40$, $T=24$, $\gamma=0.038$), office networks ($N = 219$, $n_{obs}=15$, $\gamma = 6\cdot 10^{-4}$), school networks ($N=328$, $n_{obs}=20$, $\gamma =7\cdot 10^{-5}$).
			}
			\label{fig:aucs}
		\end{center}
	\end{figure*} 
	We consider a typical risk assessment scenario of epidemic inference in which, given the network of contacts and some observations made on an epidemic realization with one initially infected individual, one has to find the probability of each individual being infected at the final time. The simulations of the epidemic realizations according to the SI model are performed using the EpiGen python package \cite{epigen} on both synthetic and real-world contact networks. The observations are performed on a random subset of the population at the final time. Once the individual probability of being infected at the final time is estimated with an epidemic inference method, the knowledge of the ground truth provided by the corresponding epidemic realization allows to compute a ROC curve of true infected individuals vs. false infected individuals. The area under the ROC curve (AUC) represents an estimate of the probability of correct classification of the individual infection states. 
	The inference through SCDC is carried out using the efficient formulation through transfer matrices, and its performance is evaluated in comparison with two well-established methods for distributed epidemic inference, the Simple Mean Field (SMF) method \cite{baker2021epidemic} and the Belief Propagation (BP) algorithm \cite{altarelli2014patient, baker2021epidemic}.
	For both BP and SCDC methods, the individual probability of being infected at the final time is computed from the corresponding total marginals once the message passing algorithm has reached convergence, i.e. the error on the cavity marginals decreases under a predefined tolerance threshold. In some cases, BP and SCDC do not reach convergence in a reasonable number of iterations (a few thousand in the case of the epidemic instances under study). The lack of convergence can be due to a relevant role played by loop structures and  long-range correlations. In such cases, the probability marginals are computed taking an average over a sufficiently large  number (up to hundreds) of iterations of the message passing update. We analyzed in greater detail how often the two algorithms converge and how convergence affects inference performance, as will be shown later in this paper. The Simple Mean Field (SMF) inference method introduced in \cite{baker2021epidemic} is instead an inference method based on the IBMF approximation for the SI dynamics in which the information provided by observations of susceptible and infected individuals is taken into account by introducing some specific constraints on the time-forward dynamics (see Ref.\cite{baker2021epidemic} for a description in the more general SIR model). 
	
    Figure \ref{fig:aucs} presents the results for various types of graphs. The top row highlights outcomes for synthetic random graphs. The first two panels (from left to right) correspond to two categories of static random networks: Watts-Strogatz graphs \cite{watts_collective_1998} and soft random geometric graphs \cite{penrose_connectivity_2016}. In the Watts-Strogatz model, the edges of a pristine network with a regular locally connected structure are rewired randomly with a probability $p_{rw}$, leading to the emergence of non-trivial small-world and clustering properties. In soft random geometric graphs, also known as proximity random graphs, individuals are distributed uniformly at random in the unit square, then only pairs at Euclidean distance $l<l_{\rm max}$ are connected with a probability which decays exponentially with $l$. Both classes of random networks are locally highly structured, with short loops and clusters. 
	The final two panels in the first row refer to synthetic contact networks generated with more realistic agent-based models, the OpenABM-Covid19 \cite{hinch_openabm_2021} and Covasim \cite{kerr_covasim_2021} models. These agent-based models are able to generate realistic contact networks on large populations, by modeling the interactions in households, schools, workplaces and other locations. Some contacts in these networks also change daily to reflect the dynamic nature of real-life interactions. Only the contact network structures generated by these agent-based models 
	over a time horizon of a few weeks is used in the present work, and the epidemic propagations are generated using the standard SI model.  
	In these networks, the link between two individuals $i$ and $j$ is assigned a weight $w_{ij}^t$, representing the aggregate duration of the contact between $i$ and $j$ in  day $t$. Given an infection rate $\gamma$ per contact per time unit, the infection probability associated to the contact is then computed as $\lambda_{ij}^t = 1-e^{-\gamma w_{ij}^t}$.
	In both cases, when a relatively small number of observations is provided at the last time, the SCDC method is able to outperform SMF and achieves accuracy on par with the BP method.
    
	The same testing framework is also employed to evaluate epidemic inference on two real contact networks, originally presented in Ref.~\cite{Genois2018}, that have been collected with RFID tags in a school (Thiers13 dataset) and in an office environment (InVS15 datasets). The contact data are collected over a period of several days, with a temporal resolution of 20 seconds, which allows for data aggregation over coarse-grained time windows of a preferred size $\tau_w$. In our study, time windows with size $\tau_w$ ranging from 3 hours to a day are considered, for a total of $T$ time steps ranging from a minimum of 12 to a maximum of 36 steps. When performing the coarse-graining procedure, the number $c_{ij}^{t}$ of contacts   between $i$ and $j$ occurring in a time window $t$ of size $\tau_w$ is computed and used to estimate the infection probability $\lambda_{ij}^t$ between the two individuals at time step $t$ as $\lambda_{ij}^t = 1-\left(1-\gamma\right)^{c_{ij}^t}$, where $\gamma$ is a common parameter describing the infectiousness of a single contact. 
	The results of epidemic risk assessment on these real-world contact networks are shown in the second row of Figure \ref{fig:aucs}, adopting the same metric used in the case of random graphs. Also in this case, for all contact networks under study, the SCDC method has a performance very close to the BP algorithm, and in general superior to the SMF heuristic. 
	
    \subsubsection{Effect of the infection probability and number of observations}

    \begin{figure}
		\begin{center}
			\includegraphics[width=\columnwidth]{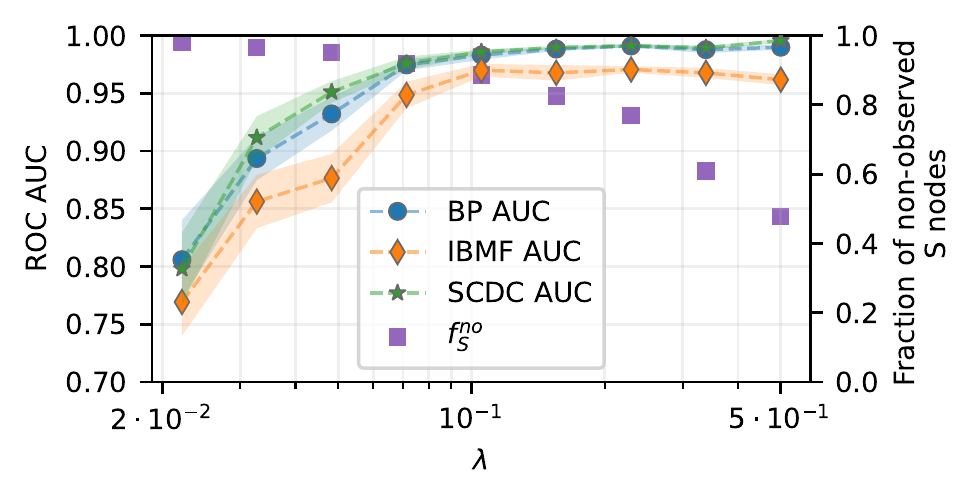}
			\caption{\textbf{Effect of $\lambda$.} AUC comparison of SCDC, BP, and SMF on soft random geometric graphs as a function of the homogeneous infection probability $\lambda$. Results are averaged over 100 instances of random networks and epidemic outbreaks with $N = 300$ nodes, $l_{\rm max} = 1.8$, $T = 15$ days, and 30\% of nodes observed at the final time. The ribbon width represents the standard error, showing the variability of the results. The figure also shows the fraction of non-observed nodes that remain susceptible at the last time, namely $f_S^{no}$.}
			\label{fig:lambda_srg}
		\end{center}
	\end{figure}
    \begin{figure}
		\begin{center}
			\includegraphics[width=\columnwidth]{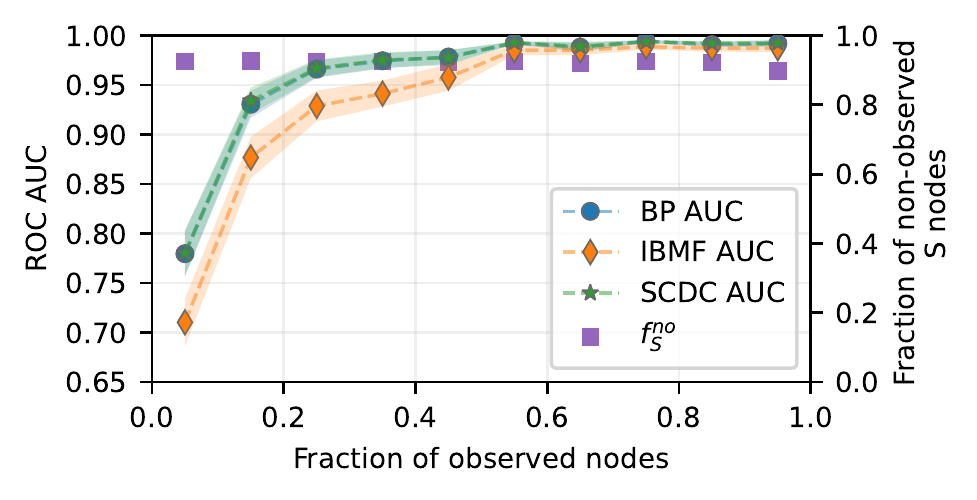}
			\caption{\textbf{Effect of the number of observations.} AUC comparison of SCDC, BP, and SMF on soft random geometric graphs as a function of the fraction of observed nodes at the last time. Results are averaged over 100 instances of random networks and epidemic outbreaks with $N = 300$ nodes, $l_{\rm max} = 1.8$, $T = 15$ days, and $\lambda=0.07$. The ribbon width represents the standard error, showing the variability of the results. The figure also shows the fraction of non-observed nodes that remain susceptible at the last time, namely $f_S^{no}$.}
			\label{fig:observations_srg}
		\end{center}
	\end{figure} 
    In addition to the networks discussed above, we further assess the performance of SCDC, BP, and IBMF on soft random geometric graphs as a function of two critical parameters: the homogeneous infection probability $\lambda$ and the fraction of observed nodes at the final time.
    
    Figure~\ref{fig:lambda_srg} illustrates the AUC achieved by the three methods as a function of $\lambda$, averaged over 100 instances of random contact networks and epidemic outbreaks. The soft random geometric graphs used in this analysis consist of $N = 300$ nodes, with a cutoff distance of $l_{\rm max} = \sqrt{1.8/N}$. Epidemic outbreaks are simulated over $T = 15$ days, with two randomly selected individuals initially infected at time $t = 0$. At the final time step, a randomly selected fraction of 30\% of the nodes is observed. The infection probability $\lambda$ is varied from 0.02 to 0.5. Surprisingly, the SCDC method shows inference performance comparable to that of BP, even for larger values of $\lambda$, despite being a small-coupling approximation to BP. In all cases, BP outperforms the IBMF method in assessing epidemic risk.

    Figure~\ref{fig:observations_srg} evaluates the effect of varying the fraction of observed nodes on the AUC. The setup is identical to that described for Figure~\ref{fig:lambda_srg}, with the infection probability fixed at $\lambda = 0.15$. The fraction of nodes observed at the final time is varied from 5\% to 95\%. As the fraction of observed nodes increases, the performance of the three methods becomes more comparable. For a smaller fraction of observed nodes, SCDC maintains performance similar to BP, while IBMF performs slightly worse. However, as the observation fraction grows, the differences between the methods diminish, and their performance becomes comparable for sufficiently large fractions of observed nodes.

    \begin{table}
    \centering
    \caption{Convergence rates of BP and SCDC for varying values of $\lambda$. The inference setting is the same of Fig.~\ref{fig:lambda_srg}.}
    \label{tab:conv_lambda}
    \begin{tabular}{c|ccccccccc}
        $\lambda$ & 0.02 & 0.03 & 0.05 & 0.07 & 0.1 & 0.15 & 0.23 & 0.34 & 0.5 \\
        \hline
        BP & 1.0 & 0.98 & 1.0 & 0.68 & 0.48 & 0.2 & 0.12 & 0.06 & 0.0 \\
        SCDC & 0.7 & 0.5 & 0.48 & 0.2 & 0.16 & 0.16 & 0.16 & 0.08 & 0.06 \\
    \end{tabular}
    \end{table}
    \begin{table}
        \centering
        \caption{Convergence rates of BP and SCDC for varying fractions of observed nodes. The inference setting is the same of Fig.~\ref{fig:observations_srg}.}
        \label{tab:conv_obs}
        \begin{tabular}{c|cccccccccc}
            F. obs. & 0.05 & 0.15 & 0.25 & 0.35 & 0.45 & 0.55 & 0.65 & 0.75 & 0.85 & 0.95 \\
            \hline
            BP & 0.8 & 0.8 & 0.8 & 0.8 & 0.8 & 0.8 & 0.8 & 0.8 & 0.8 & 0.8 \\
            SCDC & 0.8 & 0.4 & 0.22 & 0.28 & 0.26 & 0.22 & 0.38 & 0.44 & 0.4 & 0.5 \\
        \end{tabular}
    \end{table}

    In some cases, BP and SCDC do not reach convergence within a reasonable number of iterations (a few thousand in the case of the epidemic instances under study). The lack of convergence can be attributed to the relevant role played by loop structures and long-range correlations. In such cases, the probability marginals are computed by taking an average over a sufficiently large number (up to hundreds) of iterations of the message-passing updates. Notably, the instances on which the convergence of BP and SCDC is evaluated are the same as those used in Figures~\ref{fig:lambda_srg} and \ref{fig:observations_srg}. Tables~\ref{tab:conv_lambda} and \ref{tab:conv_obs} present the fraction of converged instances for both BP and SCDC under varying $\lambda$ and fractions of observed nodes, respectively. While BP consistently shows better convergence rates than SCDC, the convergence rate decreases for both methods as $\lambda$ increases. Despite this, the inference accuracy of both methods remains comparable, even when the convergence rate of SCDC is significantly lower than that of BP.

	\subsection{Inference in recurrent epidemic models}

    \begin{figure}
		\begin{center}
			\includegraphics[width=\columnwidth]{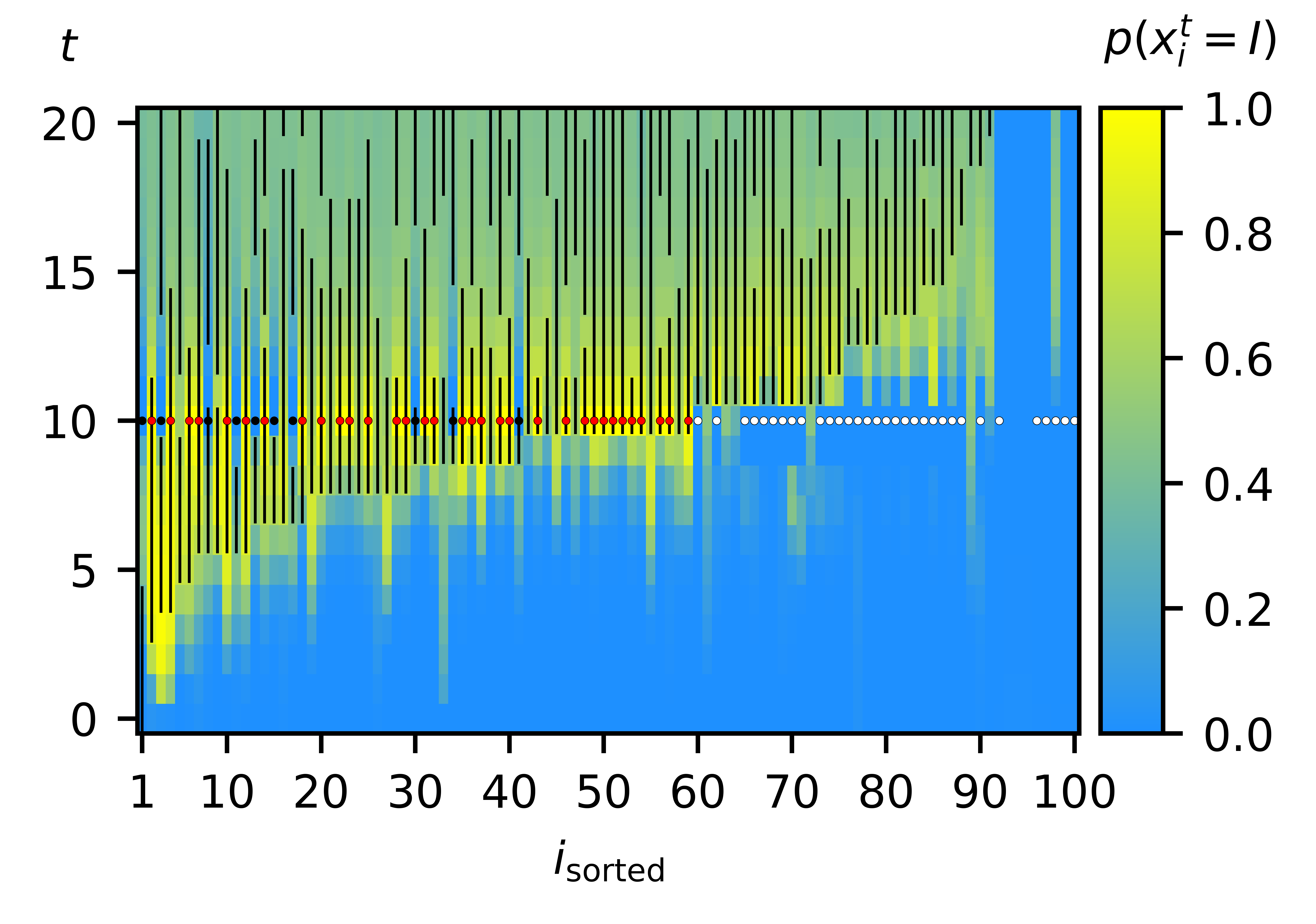}
			\caption{\textbf{Inference in recurrent epidemic models.} Posterior probability of being infected as a function of time for all nodes of an Erdős–Rényi random graph ($N = 100$ nodes and average degree $z = 3$) in a single epidemic outbreak generated according to a uniform SIRS model. The parameters of the SIRS model are $\lambda = 0.4$, $r = \sigma = 0.15$, for all nodes at all times. The same parameters were used to perform the inference. The 75\% of the nodes were randomly observed at the same time 10, and their corresponding state is represented by colored dots (white=S, red=I, black=R). Black vertical lines mark to true infection periods. Node order from left to right reproduces the order of true infection events. }
			\label{fig:SIRSinference}
		\end{center}
	\end{figure} 
	While previous results focus on the quantitative analysis of inference performances on irreversible dynamics, the present subsection aims at illustrating the potential of the SCDC method for epidemic inference on recurrent epidemic models. In order to do that, we perform a simple analysis inspired by the one already presented in recent work on Matrix Product Belief Propagation (MPBP) \cite{crottiMatrixProductBelief2023}, a novel powerful approximation method for observation-reweighted recurrent dynamics on graphs. We conducted simulations of a single epidemic outbreak using a SIRS model on an Erdős–Rényi random graph \cite{erdosRandomGraphs1959} with $N=100$ nodes and average degree $z=3$. Figure \ref{fig:SIRSinference} shows the value of the posterior marginal probability of being infected $p\left( x_i^t = 0 |  \boldsymbol{\mathcal{O}}\right)$ inferred by the SCDC method. The color scale indicating these probability values is superimposed on the black bars marking the time intervals of true infections, which enables visual inspection of the inference performance of the method.
    Notably, the SCDC method rather accurately assigns posterior marginal probabilities that closely align with the observed data, demonstrating its effectiveness even for unobserved nodes or time points that are distant from the observations. Several reinfection events are also correctly captured.

    \begin{figure}
		\begin{center}
			\includegraphics[width=\columnwidth]{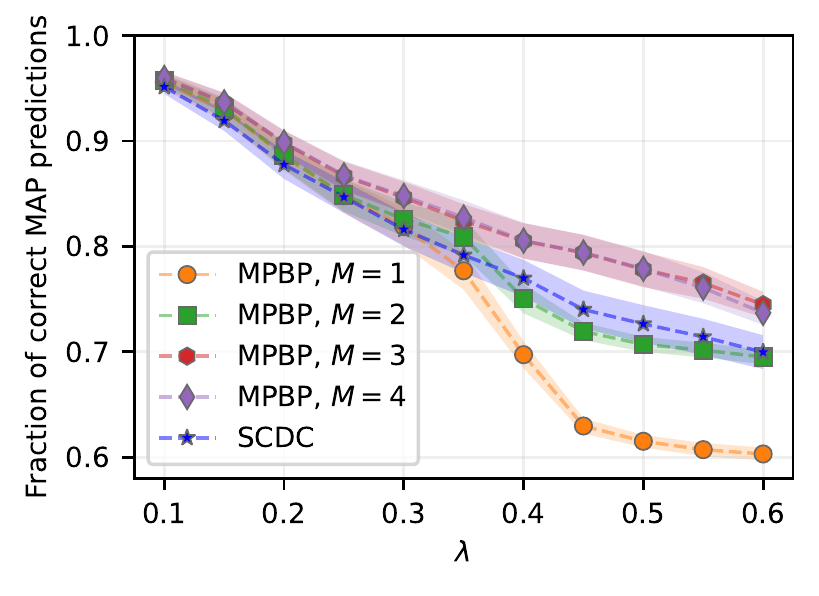}
			\caption{\textbf{Epidemic risk assessment on recurrent models.} Comparison of the inference performance of the SCDC algorithm  and the MPBP algorithm for different bond dimensions $M=1,2,3,4$ on Erdős–Rényi random graphs with $N=100$ nodes and average degree $z=2.5$. The epidemic spread follows an SIRS model with uniform infection probability $\lambda$, recovery probability $r_i^t=0.15$, and immunity-loss probability $\sigma_i^t=0.15$, for all nodes and times. The duration of the epidemic is $T=20$ steps, with an initial condition of 2 randomly chosen infected individuals at $t=0$. At $t=10$, 75\% of the nodes were randomly selected for observation. The averaged fraction of correct Maximum-a-Posteriori (MAP) predictions, computed over 50 instances of random graphs and epidemic outbreaks, is plotted as a function of $\lambda$. The ribbon width represents the standard error, showing the variability of the results.}
			\label{fig:SCDC_MPBP_comparison}
		\end{center}
	\end{figure} 
    Finally, we performed a quantitative comparison of the inference performance between the SCDC method and MPBP on recurrent epidemic processes, specifically focusing on the SIRS model. To evaluate the reconstruction accuracy of the inference algorithms, we used the fraction of correct Maximum-a-Posteriori (MAP) predictions. Let $\mathbb{P}(x_i^t)$ denote the inferred probability of node $i$ being in state $x_i^t \in \{S, I, R\}$ at time $t$, and let $X_i^t$ represent the ground truth, i.e., the true state of node $i$ at time $t$. The fraction of correct MAP predictions is defined as  
    \begin{equation}
        f_{\text{correct}}^{\text{MAP}} = \frac{1}{NT} \sum_{i=1}^N \sum_{t=1}^T \delta\left(\argmax_{x_i^t \in \{S, I, R\}} \mathbb{P}(x_i^t), X_i^t\right),
    \end{equation}
    where $\delta(x,y)$ denotes the Kronecker symbol. Figure~\ref{fig:SCDC_MPBP_comparison} compares the inference performance of the SCDC algorithm with the MPBP algorithm, evaluated at different bond dimension parameters, $M = 1, 2, 3, 4$. The bond dimension is a parameter of the MPBP method that determines the size of the matrices used in the matrix product approximation of the cavity messages. The comparison is based on SIRS epidemic models simulated on Erdős–Rényi random graphs with $N = 100$ nodes and an average degree $z = 2.5$. The recovery and immunity-loss probabilities are uniform, with $r_i^t = \sigma_i^t = 0.15$ for all nodes and time steps, while the infection probability $\lambda_i^t = \lambda$ is constant across all nodes and time steps within each simulation. For each fixed value of $\lambda$, the fraction of correct MAP predictions was averaged over 50 independent realizations of random graphs and epidemic outbreaks. At time step $t=10$, a randomly chosen fraction of 75\% of the nodes was observed, providing partial data for the inference algorithms. The plot shows the average fraction of correct MAP predictions for different values of $\lambda$.

    The results show that, for small infection probabilities $\lambda$, all algorithms perform equivalently achieving high accuracy . In this regime, the slower spread of the epidemic allows for easier inference. However, as $\lambda$ increases, the epidemic cascades become faster and more challenging to reconstruct, leading to a decline in inference performance. Notably, the SCDC algorithm performance is comparable with that of the MPBP algorithm with bond dimensions $M=2$, even in this challenging inference regime. Additionally, the MPBP algorithm does not show any improvement in inference performance for $M > 3$, as the results for $M=4$ are identical to those for $M=3$. This suggests that $M=3$ represents the maximum inference performance achievable by the MPBP algorithm in this setting.

    An important advantage of the SCDC algorithm lies in its computational efficiency: while the computational complexity of MPBP scales as $O(S^6|E|TM^6)$, the complexity of SCDC is only $O(S^2|E|T)$, where $S$ is the number of states in the compartmental epidemic model under study (e.g., $S=3$ for the SIRS model). Moreover, although MPBP, especially with sufficiently high bond dimensions $M$, achieves greater accuracy in reconstructing the epidemic spread, the SCDC algorithm provides a simple and interpretable approximation. The terms involved in the SCDC approach, such as $m$ and $\mu$, have a clear and interpretable physical meaning, making the method more accessible for further theoretical development.

    \begin{figure}
		\begin{center}
			\includegraphics[width=\columnwidth]{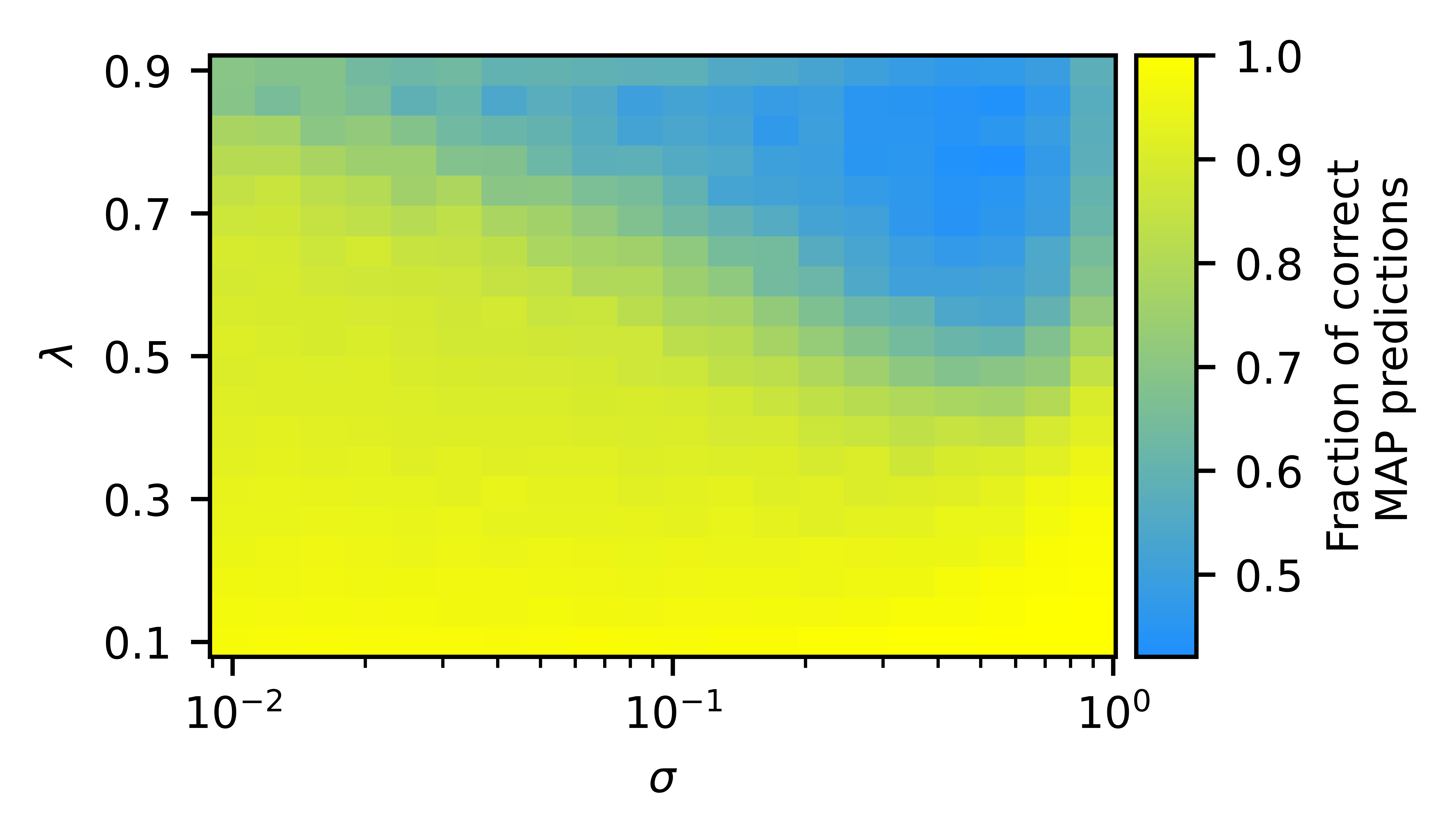}
			\caption{\textbf{Effect of $\lambda$ and $\sigma$ on epidemic reconstruction.} Fraction of correct MAP predictions as a function of the homogeneous infection probability $\lambda$ and the homogeneous recovery probability $\sigma$ for a SIS epidemic model on soft random geometric graphs with $N=300$ nodes. The cutoff distance is set to $l_{\rm max} = \sqrt{1.8/N}$. The SIS epidemic is simulated over $T=15$ days, with two randomly chosen nodes infected at time $t=0$. At time $t=7$, a randomly selected fraction of 60\% of the nodes is observed. The fraction of correct MAP predictions is averaged over 30 independent instances of random graphs and epidemic outbreaks. The color scale represents the fraction of correct MAP predictions, with higher values indicating better reconstruction accuracy. }
			\label{fig:SIS_lambda_sigma}
		\end{center}
	\end{figure} 
    As a different analysis, we now consider a SIS epidemic model instead of the previously used SIRS model. In Figure~\ref{fig:SIS_lambda_sigma}, we evaluate the fraction of correct MAP predictions as a function of both the homogeneous infection probability $\lambda$ and the homogeneous recovery probability $\sigma$. The inference is performed on soft random geometric graphs with $N=300$ nodes, where the cutoff distance is set to $l_{\rm max} = \sqrt{1.8/N}$. The SIS epidemics are simulated over $T=15$ days, with two randomly infected nodes at $t=0$. At time $t=7$, a fraction of 60\% of the nodes is observed. The fraction of correct MAP predictions was averaged over 30 independent instances of random graphs and epidemic outbreaks. The infection probability is varied from 0.1 to 0.9, while the recovery probability is varied from 0.01 to 0.9. This analysis provides a comparison of the SCDC method's performance under different epidemic dynamics, focusing on how the inference accuracy depends on the infection and recovery probabilities. The results show that the epidemic reconstruction using the SCDC method is generally accurate, except for a small region in the phase space with large values of both $\lambda$ and $\sigma$. This region corresponds to epidemic outbreaks with frequent reinfections, where inference is more challenging due to the complex dynamics of the epidemic spread.
	
	\section{Discussion}\label{sec:conclusions}
    
	The Dynamic Cavity method is a distributed technique to study discrete-state stochastic processes on graphs, which is exact on trees and often provides very good approximations on sparse graphs. While its original formulation is computationally demanding \cite{neri2009cavity,kanoria2011majority}, approximations have been introduced \cite{aurell2011message,aurell2012dynamic,del2015dynamic,aurell2017cavity,vazquez2017simple,ortega2022dynamics} and, whenever possible, more efficient parameterizations of single dynamical trajectories have been designed \cite{altarelli2013large,altarelli2014bayesian,altarelli2014patient}. In the present work, an observation-reweighted version of the dynamic cavity formulation, including individual observations, is devised to model the posterior probability of epidemic processes on contact networks. The formulation exploits a Bayesian approach and is fully equivalent to the Belief Propagation approach to epidemic trajectories \cite{altarelli2014bayesian,altarelli2014patient}. Starting from the reweighted Dynamic Cavity formulation and exploiting a small-coupling expansion, a novel set of fixed-point equations for a pair of time-dependent cavity messages $m_{i\setminus j}^t$ and $\mu_{i\setminus j}^t$ 
	is obtained. Here, $m_{i\setminus j}^t$ is the approximate probability that individual $i$ is infected at time $t$ in the cavity graph when the interaction with individual $j$ is removed, while $\mu_{i\setminus j}^t$ is a cavity field whose role depends on the presence of observations. In the absence of observations, all cavity fields $\{\mu_{i\setminus j}^t\}$ identically vanish, and  
	the dynamics, expressed solely in terms of cavity probabilities $\{m_{i\setminus j}^t\}$, becomes causal, reducing to a set of generalized mean-field equations. These time-forward equations, tested on random graphs for the SI model, yield higher accuracy compared to the commonly used individual-based mean-field equations (which they reduce to in the regime of low infectiousness and high connectivity) for all the tested settings. They are albeit less accurate than BP, which in the simplified case of non-recurrent forward dynamics coincides with the Dynamic Message Passing method \cite{lokhov2015dynamic,altarelli2013large}. Simple analyses conducted on the SI model with limited observations demonstrate that the role of the  cavity fields $\{\mu_{i\setminus j}^t\}$ is to propagate information about observations to neighboring nodes and subsequently distribute this information throughout the contact network, appropriately tilting the probabilistic weight of the associated dynamic trajectories in view of the presence of observations. The presence of observations renders the epidemic dynamics non-causal, with backward-in-time information flow, as evident from the non-uniform distribution of the normalized backward cavity messages $\tilde{\rho}_{t \leftarrow }^{i\setminus j}(x_i^t)$.  
	The main additional approximation assumed in deriving the SCDC method from the DC equations is the independence of cavity messages from the epidemic trajectory of the removed node. 
	This approximation can introduce inconsistencies with specific trajectories imposed by observations, particularly in regimes with numerous observations, including repeated observations on the same individuals. However, this issue is effectively resolved by introducing a small self-infection probability, which practically eliminated the problem in all applications considered. The SCDC algorithm shows promising effectiveness in assessing the epidemic risk of individuals, with performance comparable to that of BP, of which it is essentially an approximation, and offering improvements over other heuristic methods based on mean-field approximations. 
	As a fixed-point message passing method, a potential drawback of SCDC lies in its convergence properties. In numerical tests, SCDC experiences convergence problems similar to BP, mainly resulting from long-range correlations generated by loops in the contact graphs. Nevertheless, even in the absence of convergence, the estimated marginal probabilities often remain sufficiently accurate, enabling a reliable estimation of the epidemic risk. \\ The main advantage of the SCDC method over Dynamic Cavity and Belief Propagation for epidemic trajectories lies in its straightforward generalization to epidemic models with multiple states (e.g., SIR, SEIR) and recurrent processes (e.g., SIS, SIRS). Indeed, the fundamental components of the method and the efficient algorithm based on the temporal transfer matrix remain largely unchanged, with only modifications in the matrix dimensions and elements to accommodate the model's increased complexity. Consequently, the SCDC algorithm maintains a linear complexity with respect to the duration of the epidemic process and the number of contacts in the network. We compared SCDC with Matrix Product Belief Propagation (MPBP) \cite{crottiMatrixProductBelief2023}, a novel approximation method for observation-reweighted recurrent dynamics on graphs, based on a matrix product approximation of the cavity messages. Both methods showed comparable inference performance on SIRS models at low infection probabilities. However, as the infection probability increases, the performance gap widens in favor of MPBP. Despite this, SCDC retains a key advantage in terms of computational efficiency and offers a simpler approximation with a more intuitive physical interpretation of the involved parameters. Perhaps surprisingly, this excellent predicting power persists in regimes in which the infectiousness parameter is relatively large, outside the range that was expected due to the assumptions in the derivation, as demonstrated for the inference on SIS models. \\
	The primary limitation of the SCDC method is that its efficient formulation based on the transfer matrix is currently applicable only to Markovian models. Further study is required to develop an efficient algorithm for non-Markovian recurrent epidemic models.
	Concerning the method itself, another interesting direction for its development involves gaining a better understanding of the role of second-order terms in the small coupling expansion and developing an improved algorithm that takes them into account.
	Finally, future directions include the possibility to generalize the approach presented here to other type of dynamical processes on networks, e.g. rumor spreading processes \cite{rumordaley,rumorlatora,kanoria2011majority}.
    
	\section{Acknowledgments}
    
	Computational resources were provided by the SmartData@PoliTO \cite{smartdatapolito} interdepartmental center on Big Data and Data Science. We thank S. Crotti for suggesting the format of Figures \ref{fig:norm_mu} and \ref{fig:SIRSinference}, similar to what used in Ref.~\cite{crottiMatrixProductBelief2023}. G.C. acknowledges support by the Comunidad de Madrid and the Complutense University of Madrid (UCM) through the Atracción de Talento programs (Refs. 2019-T1/TIC-13298). F.M. acknowledges support from the project “CODE – Coupling Opinion Dynamics with Epidemics”, funded under PNRR Mission 4 "Education and Research" - Component C2 - Investment 1.1 - Next Generation EU "Fund for National Research Program and Projects of Significant National Interest" PRIN 2022 PNRR, grant code P2022AKRZ9, CUP B53D23026080001. A.B. and L.D.A acknowledge that this study was carried out within the FAIR - Future Artificial Intelligence Research project and received funding from the European Union NextGenerationEU (Piano Nazionale di Ripresa e Resilienza (PNRR)–Missione 4 Componente 2, Investimento 1.3–D.D. 1555 11/10/2022, PE00000013). This manuscript reflects only the authors’ views and opinions, neither the European Union nor the European Commission can be considered responsible for them.

%%%%%%%%%%%%%%%%%%%%%%%%%%%%%%%%%%%%%%%%%%%%%%%%%%%%%%%%%%%%%%%%%%%%%%%%%%%%%%%%%%
    \appendix

    \section{Belief propagation equations for the SI model}\label{app:BP_derivation}
    \begin{figure}
      \begin{center}
      \includegraphics{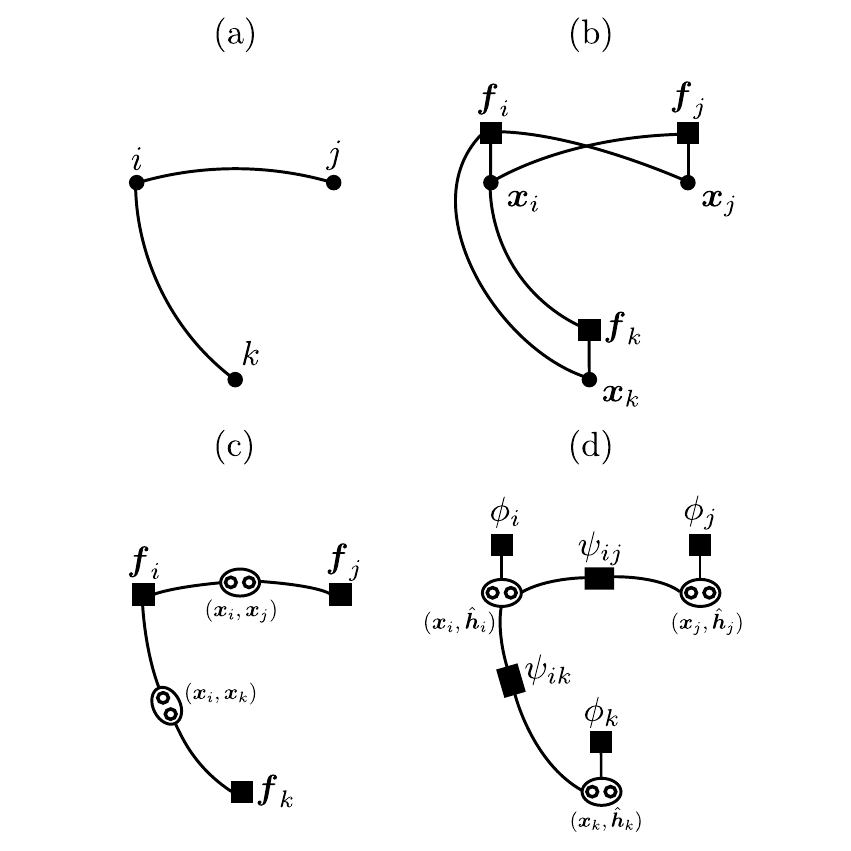}
      \caption{{\bf Factor graph representation for epidemic inference.} (a) Original contact graph. (b) Loopy, naive factor graph associated to the graphical model Eq.~\eqref{eq:factorized_prob}. (c) Disentangled dual factor graph for the graphical model Eq.~\eqref{eq:factorized_prob}. The factor graph maintain the structure of the original contact graph. (c) Disentangled factor graph for the graphical model interpretation of the dynamical partition function in Eq.~\eqref{eq:dynamicalZforFG} with factors as given in Eqs.~\eqref{eq:Factor_phi}-\eqref{eq:Factor_psi}.}
      \label{fig:factorgraph}
      \end{center}
    \end{figure}
    \begin{figure*}
      \centering
      \includegraphics{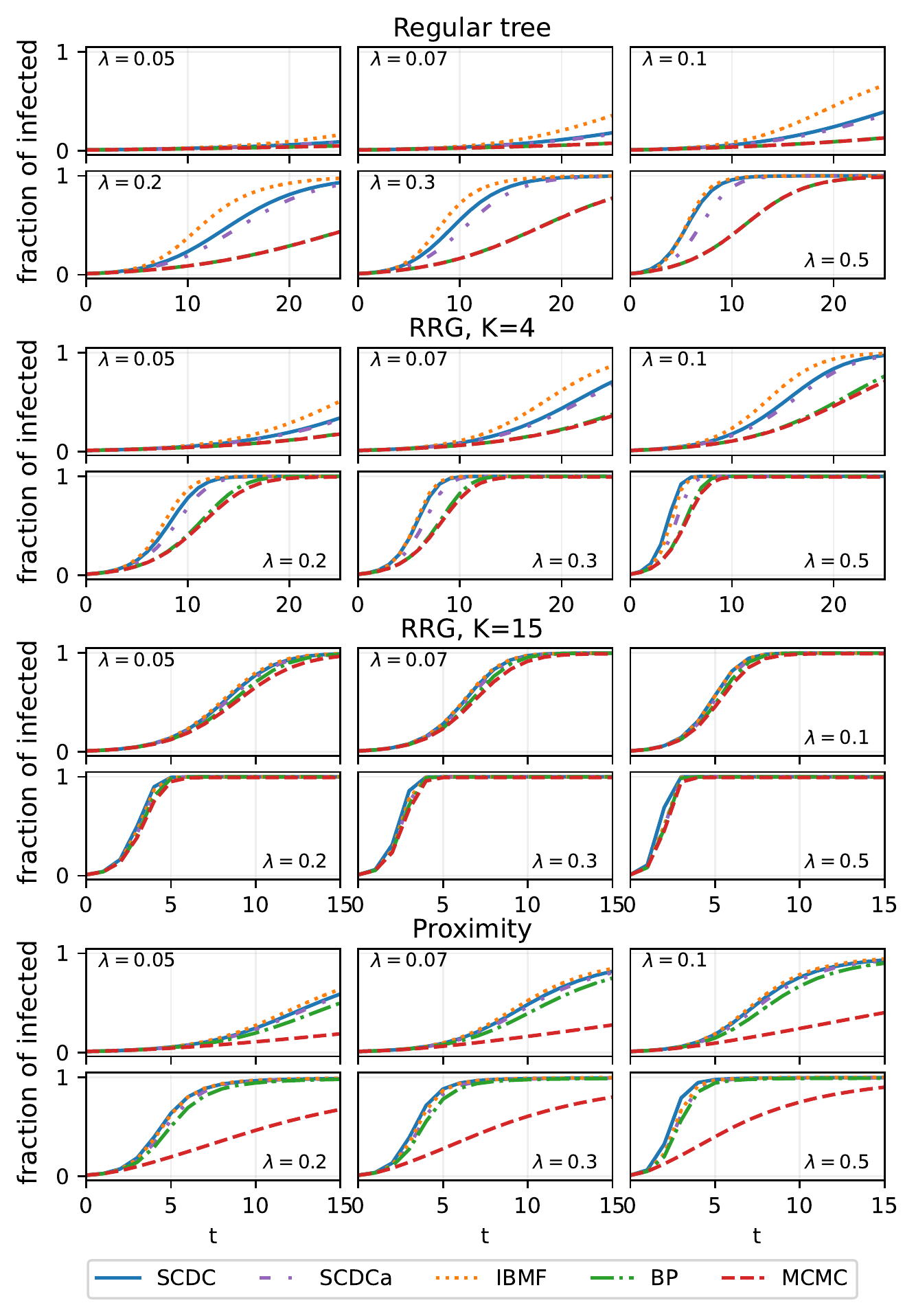}
      \caption{\textbf{Forward dynamics with varying $\lambda$}. Fraction of infected individuals against epidemic time, with four different static contact graphs. From top to bottom:  regular tree with degree $K=4$ and $N=485$, RRG with $N=500$ and degree $K=4$, RRG with $N=500$ and $K=15$, proximity graph with $N=500$. Comparison is shown between SCDC, IBMF, BP and Monte Carlo simulations (with $10^4$ samples). All the links have the same infection probability $\lambda$, whose value is reported inside each panel. Six different values of $\lambda$ are used. In all cases, the probability of each individual being infected at time $t=0$ is set to $p_0(x_i^0)=5\slash N$, and the self-infection $\varepsilon_i^t$ is set to $0$ for every node $i$.}
      \label{fig:varlambda_fwd}
	  \end{figure*}
    The posterior probability in Eq.~\eqref{eq:posterior2} can be rewritten in the following factorized form:  
    \begin{equation}\label{eq:factorized_prob}
        p({\bm X}|{\bm O})\propto \prod_{t=0}^{T-1}\prod_{i=1}^N f_i^{t+1}(x_i^{t+1},{\bm x}_{\partial i}^t,x_i^t|\boldsymbol{\mathcal{O}}),
    \end{equation}  
    where the factors \( f_i^{t+1} \) encode both the local transition dynamics and the influence of observations, and are defined as:  
    \begin{align}
        f_i^{t+1}(x_i^{t+1},{\bm x}_{\partial i}^t,x_i^t|\boldsymbol{\mathcal{O}}) & =(p_0(x_i^0)p(O_i^0|x_i^0))^{\delta_{t,0}} \nonumber \\
        &\quad \times W_i(x_i^{t+1}|{\bm x}^t)p(O_i^t|x_i^t).
    \end{align}  
    Here, \( W_i(x_i^{t+1}|{\bm x}^t) \) represents the transition rate from \( x_i^t \) to \( x_i^{t+1} \), which depends solely on the state of the neighbors at time \( t \), denoted as \( {\bm x}_{\partial i}^t \). Although the factor graph associated with this posterior distribution exhibits a substantial number of short loops, the structure can be disentangled by leveraging the interactions between trajectories of neighboring nodes. Specifically, pairs of trajectories  $({\bm x}_i, {\bm x}_j)$ that are connected by an edge in the original contact graph can be grouped together. This procedure leads to a reformulation in terms of a new graphical representation, known as the \textit{dual factor graph}.  
    
    In the dual factor graph, the variable nodes correspond to pairs of trajectories $({\bm x}_i, {\bm x}_j)$, which are associated with the edges of the original contact graph. The factor nodes, on the other hand, correspond to the vertices  $i$ of the contact graph. A visual depiction of this transformation is shown in Figure~\ref{fig:factorgraph}, where panel (b) illustrates the original loopy factor graph, and panel (c) demonstrates the disentangled dual factor graph. The factors in the dual representation are obtained by grouping the contributions over all time steps for each vertex, i.e. ${\bm f}_i({\bm x}_i, \{\bm x_k\}_{k\in\partial i}|\boldsymbol{\mathcal{O}})=\prod_{t=0}^{T-1}f_i^{t+1}(x_i^{t+1},{\bm x}_{\partial i},x_i^t|\boldsymbol{\mathcal{O}})$. A more comprehensive discussion of this step, including its theoretical implications and practical benefits, can be found in Refs.~\cite{altarelli2013large, altarelliOptimizingSpreadDynamics2013b, altarelli2014bayesian, altarelli2014patient, baker2021epidemic}.

    The BP equations on the dual factor graph are \cite{crottiMatrixProductBelief2023}
    \begin{subequations}
    \begin{align}
        \hat{c}_{i\setminus j}({\bm x}_i,{\bm x}_j|\boldsymbol{\mathcal{O}}) & \propto \sum_{{\bm x}_{\partial i \setminus j}}\prod_{t=0}^{T-1}f_i^{t+1}(x_i^{t+1},{\bm x}_{\partial i}^t,x_i^t|\boldsymbol{\mathcal{O}})\nonumber\\
        & \quad \times \prod_{k\in\partial i\setminus j} \hat{c}_{k\setminus i}({\bm x}_k,{\bm x}_i|\boldsymbol{\mathcal{O}})\\
        & \propto p_0(x_i^0)p({\bm O}_i|{\bm x}_i)  \sum_{{\bm x}_{\partial i \setminus j}} \prod_{t=0}^{T-1}  W_i(x_i^{t+1}|{\bm x}^t)\nonumber\\
        & \quad \times \prod_{k\in\partial i\setminus j} \hat{c}_{k\setminus i}({\bm x}_k,{\bm x}_i|\boldsymbol{\mathcal{O}}).
    \end{align}
    \end{subequations}
    The explicit form of the transition rate is
    \begin{subequations}
    \begin{align}
         W_i(x_i^{t+1}|{\bm x}^t) & = \delta_{x_i^{t+1},S}\delta_{x_i^{t},S}\alpha_i^te^{\sum_{k\in\partial i}\nu_{ki}^{t}\delta_{x_{k}^{t},I}}\nonumber\\
         & \quad + \delta_{x_i^{t+1},I}(1-\delta_{x_i^{t},S}\alpha_i^te^{\sum_{k\in\partial i}\nu_{ki}^{t}\delta_{x_{k}^{t},I}})\\
         & = \alpha_i^te^{\sum_{k\in\partial i}\nu_{ki}^{t}\delta_{x_{k}^{t},I}} (\delta_{x_i^{t+1},x_i^t}-\delta_{x_i^{t+1},I})\nonumber \\
         & \quad +\delta_{x_i^{t+1},I}.
    \end{align}
    \end{subequations}
    The dynamic cavity equations Eq.\eqref{eq:dyncav} are therefore obtained from the BP equations with the simple change of variables $s_i^t=\nu_{ji}^t\delta_{x_j^t,I}$. The local fields $s_i^t$ act as proxies for the missing variables $x_j^t$ of the dual factor graph. 
    
    \section{Path integral derivation of the dynamic cavity equations for the SI model}\label{app:derivation_dynamic_cavity}
    
    In this Section, we present a detailed derivation of the dynamic cavity equations, Eqs.~\eqref{eq:dyncav} and \eqref{eq:dyncav_fourier}, which is equivalent to the one proposed in Appendix~\ref{app:BP_derivation}. This derivation leverages a path-integral representation of the stochastic epidemic dynamics of the SI model. The approach is based on interpreting the (Markovian) update rules governing the discrete-time stochastic process as a set of dynamical constraints imposed on the degrees of freedom of the system, specifically the binary variables $\{x_i^t\}$. These variables describe the state of each node $i$ at time $t$, with $x_i^t \in \{S, I\}$ indicating whether the node is susceptible or infected, respectively. The derivation is further grounded in the definition of a dynamic partition function, which encapsulates the stochastic dynamics of the system and incorporates the interplay between the microscopic states of the nodes and their interactions over time. 

    The dynamic partition function is defined as follows:
    \begin{align}
    \mathcal{Z}(\boldsymbol{\mathcal{O}}) = \sum_{{\bm X}} p\left( {\bm X}|\boldsymbol{\mathcal{O}}\right) &\propto \sum_{{\bm X}}\quad\prod_{i=1}^{N}  p_0(x_i^0) p({\bm O}_i|{\bm x}_i)\nonumber\\ & \quad \times  \prod_{t=0}^{T-1}W_i(x_i^{t+1}|{\bm x}^t).\label{eq:dynPartitionFunction}
    \end{align}
    To proceed, we explicitly express $W_i(x_i^{t+1}|{\bm x}^t)$ in terms of the possible transitions:
    \begin{equation}
    W_i(x_i^{t+1}|{\bm x}^t) \propto \delta_{x_i^{t+1},S}W_i(S|{\bm x}^t) + \delta_{x_i^{t+1},I}W_i(I|{\bm x}^t),
    \end{equation}
    where $\delta_{x_i^{t+1},S}$ and $\delta_{x_i^{t+1},I}$ are Kronecker delta functions ensuring that only the appropriate transition probabilities contribute to the dynamics. Substituting this into the partition function, we have
    \begin{widetext}
    \begin{equation}
    \mathcal{Z}(\boldsymbol{\mathcal{O}}) \propto \sum_{{\bm X}} \prod_{i=1}^{N} p_0(x_i^0) p({\bm O}_i|{\bm x}_i) \prod_{t=0}^{T-1} \left[\delta_{x_i^{t+1},S}W_i(S|{\bm x}^t) + \delta_{x_i^{t+1},I}W_i(I|{\bm x}^t)\right].
    \end{equation}
    The transition probabilities $W_i(x_i^{t+1}|{\bm x}^t)$ depend on the interactions of node $i$ with its neighbors, which are captured through a local external field, $h_i^t$. Specifically, the field is defined as:
    \begin{equation}
    h_i^t = \sum_j \nu_{ji}^t \delta_{x_j^t,I},
    \end{equation}
    where $\nu_{ji}^t$ represents the interaction strength between nodes $j$ and $i$ at time $t$. The dependence of $h_i^t$ on the states of neighboring nodes is enforced by introducing a Dirac delta function $
    \delta(h_i^t - \sum_j \nu_{ji}^t \delta_{x_j^t,I})$. By substituting this definition of the local field into the dynamic partition function, we obtain
    \begin{equation}
    \mathcal{Z}(\boldsymbol{\mathcal{O}}) \propto \sum_{{\bm X}} \prod_{i=1}^{N} p_0(x_i^0) p({\bm O}_i|{\bm x}_i) \prod_{t=0}^{T-1} \int_{-\infty}^\infty dh_i^t \left[\delta_{x_i^{t+1},S}\delta_{x_i^t,S}\alpha_i^t e^{h_i^t} + \delta_{x_i^{t+1},I}\left(1 - \delta_{x_i^t,S}\alpha_i^t e^{h_i^t}\right)\right] \delta\left(h_i^t - \sum_j \nu_{ji}^t \delta_{x_j^t,I}\right),
    \end{equation}
    where $\alpha_i^t$ encodes the self-infection probability, as defined in Sec.~\ref{sec:stochmodel}. 
    
    To facilitate computation, we use the integral representation of the Dirac delta function:
    \begin{equation}
    \delta(h) = \int_{-\infty}^\infty \frac{d\hat{h}}{2\pi} e^{{\rm i}\hat{h} h},
    \end{equation}
    allowing us to introduce an auxiliary integration variable $\hat{h}_i^t$ for each $h_i^t$. Substituting this representation into the partition function and integrating out the $h_i^t$ variables introduces new interaction terms in the exponential form. These terms reflect the contributions of the external fields from neighbors and form the basis for deriving the dynamic cavity equations.
    \begin{subequations}
    \begin{align}
    \mathcal{Z}(\boldsymbol{\mathcal{O}}) & \propto \sum_{{\bm X}} \prod_{i=1}^{N}  p_0(x_i^0) p({\bm O}_i|{\bm x}_i) \prod_{t=0}^{T-1}\int_{-\infty}^\infty d h_i^t \left[\alpha_i^te^{h_i^t}\delta_{x_i^t,S}(\delta_{x_i^{t+1},S}-\delta_{x_i^{t+1},I})+\delta_{x_i^{t+1},I}\right] \int_{-\infty}^\infty \frac{d\hat{h}_i^t}{2\pi} e^{{\rm i}\hat{h}_i^t\left(h_i^t-\sum_j\nu_{ji}^t\delta_{x_j^t,I}\right)}\\
     & \propto \sum_{{\bm X}} \prod_{i=1}^{N}  p_0(x_i^0) p({\bm O}_i|{\bm x}_i) \prod_{t=0}^{T-1}\int_{-\infty}^\infty d \hat{h}_i^t \left[\delta(\hat{h}_i^t-{\rm i})\alpha_i^te^{h_i^t}\delta_{x_i^t,S}(\delta_{x_i^{t+1},S}-\delta_{x_i^{t+1},I})+\delta(\hat{h}_i^t)\delta_{x_i^{t+1},I}\right] e^{-{\rm i}\hat{h}_i^t\sum_j\nu_{ji}^t\delta_{x_j^t,I}}\\
     & \propto \sum_{{\bm X}} \prod_{i=1}^{N}  p_0(x_i^0) p({\bm O}_i|{\bm x}_i) \prod_{t=0}^{T-1}\int_{-\infty}^\infty d \hat{h}_i^t \left[\delta(\hat{h}_i^t-{\rm i})\alpha_i^te^{h_i^t}(\delta_{x_i^{t+1},x_i^t}-\delta_{x_i^{t+1},I})+\delta(\hat{h}_i^t)\delta_{x_i^{t+1},I}\right] e^{-{\rm i}\hat{h}_i^t\sum_j\nu_{ji}^t\delta_{x_j^t,I}}\\
     & \propto \sum_{{\bm X}} \int{D}\hat{{\bm H}} \prod_{i=1}^{N}  p_0(x_i^0) p({\bm O}_i|{\bm x}_i) \prod_{t=0}^{T-1} \left[\delta(\hat{h}_i^t-{\rm i})\alpha_i^te^{h_i^t}(\delta_{x_i^{t+1},x_i^t}-\delta_{x_i^{t+1},I})+\delta(\hat{h}_i^t)\delta_{x_i^{t+1},I}\right] e^{-{\rm i}\hat{h}_i^t\sum_j\nu_{ji}^t\delta_{x_j^t,I}}\\
     &\propto \sum_{{\bm X}} \int{D}\hat{{\bm H}} \prod_{i=1}^{N}  p_0(x_i^0) p({\bm O}_i|{\bm x}_i) \prod_{t=0}^{T-1} \left[\delta(\hat{h}_i^t-{\rm i})\alpha_i^te^{h_i^t}(\delta_{x_i^{t+1},x_i^t}-\delta_{x_i^{t+1},I})+\delta(\hat{h}_i^t)\delta_{x_i^{t+1},I}\right] \nonumber\\
     &\quad \times \prod_{j>i} e^{-{\rm i}\left(\hat{h}_i^t\nu_{ji}^t\delta_{x_j^t,I}+\hat{h}_j^t\nu_{ij}^t\delta_{x_i^t,I}\right)},
    \end{align}
    \end{subequations}
    \end{widetext}
    where $\int{D}\hat{{\bm H}} =  \prod_{i=1}^N \prod_{t=0}^{T-1} (\int_{-\infty}^{+\infty}d\hat{h}_i^t)$ for shortness of notation. We can simplify the notation by introducing the local non-interacting and interacting actions Eqs.~\eqref{eq:local_nonint_action} and \eqref{eq:local_int_action}. The dynamical partition function simplifies to
    \begin{equation}\label{eq:dynamicalZforFG}
        \mathcal{Z}(\boldsymbol{\mathcal{O}})\propto \sum_{{\bm X}} \int{D}\hat{{\bm H}} \prod_{i=1}^{N}  p_0(x_i^0) p({\bm O}_i|{\bm x}_i) e^{S_i^0}\prod_{j>i} e^{S_{ij}^{int}},
    \end{equation}
    where $S_i^0$ depends on the local variable-conjugate field trajectory $({\bm x}_i,\hat{\bm h}_i)$ and $S_{ij}^{int}$ depends on the two local variable-conjugate field trajectories $({\bm x}_i,\hat{\bm h}_i),\,({\bm x}_j,\hat{\bm h}_j)$.
    The probabilistic weight associated with the dynamic partition function $\mathcal{Z}(\boldsymbol{\mathcal{O}})$ is now in a form that can be represented as a graphical model, in which the variable nodes correspond to the spatio-temporal variables ${\bm x}_i$ and $\hat{{\bm h}}_i$ and there are two types of factor nodes (see Figure~\ref{fig:factorgraph}): single-node factors 
    \begin{equation}\label{eq:Factor_phi}
    \phi_i(({\bm x}_i,\hat{{\bm h}}_i)) =  p_0(x_i^0) p({\bm O}_i|{\bm x}_i) e^{S_i^0},
    \end{equation}
    and factors involving pairs of variables on neighboring nodes at the same time 
    \begin{equation}\label{eq:Factor_psi}
    \psi_{ij}(({\bm x}_i ,\hat{{\bm h}}_i),({\bm x}_j,\hat{{\bm h}}_j)) =  e^{S_{ij}^{int}} .
    \end{equation} 
    By grouping together single-node variables at all times, i.e. trajectories $({\bm x}_i,\hat{{\bm h}}_i)= ((x_i^0,\dots, x_i^T),(\hat{h}_i^0,\dots, \hat{h}_i^T))$, the resulting factor graph reproduces the topology of the underlying interaction graph.
    It should be noted that the choice of variable grouping in this approach disentangles the locally-loopy structure of the factor graph associated with the space-time problem. This disentanglement is achieved due to the linear coupling between variables on neighboring nodes, which is obtained by introducing auxiliary local fields   $\hat{{\bm h}}_i$.
    
    According to this graphical model construction, we obtain the {\it dynamic cavity equations} Eq.~\eqref{eq:dyncav_fourier} as an ansatz for describing the stochastic dynamics associated with the dynamic partition function $\mathcal{Z}(\boldsymbol{\mathcal{O}})$ on a tree-like interaction graph,
    \begin{align}
        c_{i\setminus j}({\bm x}_{i},\hat{{\bm h}}_{i} |  \boldsymbol{\mathcal{O}}) &  = \frac{p_0\left(x_{i}^{0}\right)}{\mathcal{Z}_{i\setminus j}(\boldsymbol{\mathcal{O}})} p\left( {\bm O}_i |  {\bm x}_i \right) e^{S_i^0}\nonumber \\
        & \times \prod_{k\in\partial i\setminus j} \sum_{{\bm x}_{k}}\int D\hat{{\bm h}}_{k}\, c_{k\setminus i}({\bm x}_{k},\hat{{\bm h}}_{k} |  \boldsymbol{\mathcal{O}})e^{S_{ik}^{int}}.
    \end{align}
    Then, using the Fourier transforms defined as Eqs.~\eqref{eq:fourier_transf} the dynamic cavity equations can be written as
    \begin{widetext}
    \begin{subequations}
    \begin{align}
        \frac{\hat{c}_{i\setminus j}({\bm x}_{i},{\bm s}_{i}|\boldsymbol{\mathcal{O}})}{p_0(x_{i}^{0}) p\left( {\bm O}_i |  {\bm x}_i \right)} & \propto  \int D\hat{{\bm h}}_i \left(\prod_{t} e^{-{\rm i} s_i^t \hat{h}_i^t}\right) e^{S_i^0} \prod_{k\in\partial i\setminus j} \sum_{{\bm x}_{k}}\int D\hat{{\bm h}}_{k} \int D{\bm s}_k \left(\prod_{t}\frac{e^{{\rm i} s_k^t \hat{h}_k^t}}{2\pi}\right)\hat{c}_{k\setminus i}({\bm x}_{k},\hat{{\bm h}}_{k} | \boldsymbol{\mathcal{O}}) e^{S_{ik}^{int}} \\
         & \propto  \prod_{t} \int d\hat{h}_i^t \, e^{-{\rm i} s_i^t \hat{h}_i^t} \left[\delta(\hat{h}_i^t-{\rm i})\alpha_i^te^{h_i^t}(\delta_{x_i^{t+1},x_i^t}-\delta_{x_i^{t+1},I})+\delta(\hat{h}_i^t)\delta_{x_i^{t+1},I}\right] \nonumber \\
         & \quad \times \prod_{k\in\partial i\setminus j} \left( \prod_t \sum_{{x}_{k}^t} \int d\hat{h}_k^t ds_k^t \frac{e^{{\rm i} s_k^t \hat{h}_k^t}}{2\pi}e^{-{\rm i}\left(\hat{h}_k^t\nu_{ik}^t\delta_{x_i^t,I}+\hat{h}_i^t\nu_{ki}^t\delta_{x_k^t,I}\right)}\right)\hat{c}_{k\setminus i}({\bm x}_{k},{\bm s}_{k} | \boldsymbol{\mathcal{O}}).
    \end{align}
    \end{subequations}
    The expression can be simplified by performing the integrals over the conjugate fields coming from the neighbors $\hat{{h}}_k^t$ first, then over the neighboring fields $s_k^t$,
    \begin{subequations}
    \begin{align}
        \frac{\hat{c}_{i\setminus j}({\bm x}_{i},{\bm s}_{i}|\boldsymbol{\mathcal{O}})}{p_0(x_{i}^{0}) p\left( {\bm O}_i |  {\bm x}_i \right)} & \propto \prod_{t} \int d\hat{h}_i^t \, e^{-{\rm i} s_i^t \hat{h}_i^t} \left[\delta(\hat{h}_i^t-{\rm i})\alpha_i^te^{h_i^t}(\delta_{x_i^{t+1},x_i^t}-\delta_{x_i^{t+1},I})+\delta(\hat{h}_i^t)\delta_{x_i^{t+1},I}\right] \nonumber \\
         & \quad \times \prod_{k\in\partial i\setminus j} \left( \prod_t \sum_{{x}_{k}^t} \int ds_k^t \, \delta(s_k^t-\nu_{ik}^t    \delta_{x_i^t,I}) e^{-{\rm i}\hat{h}_i^t\nu_{ki}^t\delta_{x_k^t,I}}\right)\hat{c}_{k\setminus i}({\bm x}_{k},{\bm s}_{k} | \boldsymbol{\mathcal{O}})\\
         & \propto \prod_{t} \int d\hat{h}_i^t \, e^{-{\rm i} s_i^t \hat{h}_i^t} \left[\delta(\hat{h}_i^t-{\rm i})\alpha_i^te^{h_i^t}(\delta_{x_i^{t+1},x_i^t}-\delta_{x_i^{t+1},I})+\delta(\hat{h}_i^t)\delta_{x_i^{t+1},I}\right] \nonumber \\
         & \quad \times \prod_{k\in\partial i\setminus j} \left( \prod_t \sum_{{x}_{k}^t} e^{-{\rm i}\hat{h}_i^t\nu_{ki}^t\delta_{x_k^t,I}}\right)\hat{c}_{k\setminus i}({\bm x}_{k},{\bm \nu}_{ik} {\bm x}_{i} | \boldsymbol{\mathcal{O}})\\
         & \propto \sum_{{\bm x}_{\partial i \setminus j}}\left(\prod_{k\in\partial i\setminus j}\hat{c}_{k\setminus i}({\bm x}_{k},{\bm \nu}_{ik} {\bm x}_{i} | \boldsymbol{\mathcal{O}})\right)  \prod_t \int d\hat{h}_i^t \, e^{-{\rm i} \hat{h}_i^t \left(s_i^t + \sum_{k\in \partial i\setminus j} \nu_{ki}^t\delta_{x_k^t,I}\right)}\nonumber \\
         & \quad \times \left[\delta(\hat{h}_i^t-{\rm i})\alpha_i^te^{h_i^t}(\delta_{x_i^{t+1},x_i^t}-\delta_{x_i^{t+1},I})+\delta(\hat{h}_i^t)\delta_{x_i^{t+1},I}\right],
    \end{align}
    \end{subequations}
    and finally over the local conjugate fields $\hat{{h}}_i^t$
    \begin{equation}
        \frac{\hat{c}_{i\setminus j}({\bm x}_{i},{\bm s}_{i}|\boldsymbol{\mathcal{O}})}{p_0(x_{i}^{0}) p\left( {\bm O}_i |  {\bm x}_i \right)} \propto \sum_{{\bm x}_{\partial i \setminus j}}\hat{c}_{k\setminus i}({\bm x}_{k},{\bm \nu}_{ik} {\bm x}_{i} | \boldsymbol{\mathcal{O}}) \prod_t \left[ \delta_{x_i^{t+1},I} + e^{s_i^t + \sum_{k\in \partial i\setminus j} \nu_{ki}^t\delta_{x_k^t,I}} \alpha_i^te^{h_i^t}(\delta_{x_i^{t+1},x_i^t}-\delta_{x_i^{t+1},I}) \right].
    \end{equation}
    \end{widetext}
    By introducing the dynamic cavity action Eq.~\eqref{eq:cav_dyn_action} we obtain the dynamic cavity equations Eq.~\eqref{eq:dyncav}, where cavity messages are now a function of the variable-field trajectories ${\bm x}_i,{\bm s}_i$
    \begin{align}
        \hat{c}_{i\setminus j}\left({\bm x}_{i},{\bm s}_{i}|  \boldsymbol{\mathcal{O}} \right) &  \propto p_0(x_{i}^{0}) p\left( {\bm O}_i |  {\bm x}_i \right) \sum_{{\bm x}_{\partial i \setminus j}} e^{\hat{S}_{i\setminus j}}  \nonumber \\
        &\quad \times \prod_{k\in\partial i\setminus j}  \hat{c}_{k\setminus i}\left({\bm x}_{k},{\bm \nu}_{ik} {\bm x}_{i}| \boldsymbol{\mathcal{O}}  \right).
    \end{align}    
    
    \section{Reduction to the time-forward equations in the absence of observations}\label{app:derivationtimeforward}
    
    A major consequence of the introduction of time-forward messages $\rho_{\rightarrow t}^{i\setminus j}$ and time-backward messages $\rho_{t\leftarrow}^{i\setminus j}$ is that, in the absence of observations, it is possible to prove that the quantities $\mu_{i\setminus j}^{t}$ have to vanish for all edges $(i,j)$ and times $t$ and then recover a purely time-forward dynamics.
    
    Using the definition of $m_{i\setminus j}^t$ by means of the quantities $\rho_{\rightarrow t}^{i\setminus j}\left(x_i^t\right)$ and $\rho_{t\leftarrow}^{i\setminus j}\left(x_i^t\right)$, i.e. Eq~\eqref{eq:mij_rho}, but performing the slicing one time step later, we obtain 
    \begin{subequations}
    \begin{align}
    m_{i\setminus j}^{t} & = \frac{1}{Z_{i\setminus j}^t} \sum_{x_{i}^{t},x_{i}^{t+1}}\rho_{\rightarrow t}^{i\setminus j}\left(x_{i}^{t}\right)\delta_{x_{i}^{t},I}M_{x_{i}^{t}x_{i}^{t+1}}^{i\setminus j}\rho_{t+1\leftarrow}^{i\setminus j}\left(x_{i}^{t+1}\right)\\
     & =\frac{1}{Z_{i\setminus j}^t} \rho_{\rightarrow t}^{i\setminus j}\left(I\right)M_{t, II}^{i \setminus j}\rho_{t+1\leftarrow}^{i\setminus j}\left(I\right)\label{eq:next_slice_m}
    \end{align}
    \end{subequations}
    or slicing one time step earlier,
    \begin{subequations}
    \begin{align}
    m_{i\setminus j}^{t} & = \frac{1}{Z_{i\setminus j}^t} \sum_{x_{i}^{t-1},x_{i}^{t}}\rho_{\rightarrow t-1}^{i\setminus j}\left(x_{i}^{t-1}\right)x_{i}^{t}M_{x_{i}^{t-1}x_{i}^{t}}^{i\setminus j}\rho_{t\leftarrow}^{i\setminus j}\left(x_{i}^{t}\right)\\
    & = \frac{1}{Z_{i\setminus j}^t} \left[\rho_{\rightarrow t-1}^{i\setminus j}\left(I\right)M_{t-1, II}^{i \setminus j} \rho_{t\leftarrow}^{i\setminus j}\left(I\right)\right. \nonumber\\
    &\quad \left. + \rho_{\rightarrow t-1}^{i\setminus j}\left(S\right)M_{t-1, SI}^{i \setminus j} \rho_{t\leftarrow}^{i\setminus j}\left(I\right) \right],
    \end{align}
    \end{subequations}
    where we have introduced the normalization constant $Z_{i\setminus j}^t=\sum_{x_i^t}\rho_{\rightarrow t}^{i\setminus j}(x_i^t)\rho_{t\leftarrow }^{i\setminus j}(x_i^t)$. It is straightforward to show that it corresponds to the normalization constant $\mathcal{Z}_{i\setminus j}$, defined in Eqs.~\eqref{eq:noncausalMFzij} and \eqref{eq:partition_function}. It follows that the normalization is independent of time. Using Eq.~\eqref{eq:next_slice_m} it is possible to express $m_{i\setminus j}^t$ as function of  $m_{i\setminus j}^{t-1}$
    \begin{subequations}
    \begin{align}
    m_{i\setminus j}^{t} & =  m_{i\setminus j}^{t-1} + \frac{\rho_{\rightarrow t-1}^{i\setminus j}\left(S\right)M_{t-1, SI}^{i \setminus j} \rho_{t\leftarrow}^{i\setminus j}\left(1\right)}{\mathcal{Z}_{i\setminus j}}    \\ 
    & = m_{i\setminus j}^{t-1} +\left(1-m_{i\setminus j}^{t-1}\right)\nonumber\\
    & \quad\times \frac{ \left(1 - g_{i\setminus j}^{t-1}\right) \rho_{t\leftarrow}^{i\setminus j}\left(I\right)}{ \left(1 - g_{i\setminus j}^{t-1}\right)\rho_{t\leftarrow}^{i\setminus j}\left(I\right) + g_{i\setminus j}^{t-1} \rho_{t\leftarrow}^{i\setminus j}\left(S\right)}. \label{eq:almostforward}
    \end{align}
    \end{subequations}
    As already stressed in the main text, the last expression does not represent a time-forward equation because the quantities $\rho_{t\leftarrow}^{i\setminus j}(x_i^t)$ are computed backward in time from $T$ to step $t$. Time-forward dynamics is recovered if the two time-backward messages are uniform, which is expected to occur in the absence of observations at later times. To prove this, one can first notice that from \eqref{eq:muij_rho},  $\mu_{i\setminus j}^{t}=0$ if the time-backward messages are uniform, i.e. if $\rho_{t+1\leftarrow}^{i\setminus j}(S) = \rho_{t+1\leftarrow}^{i\setminus j}(I)$. In the absence of observations also the inverse implication is true: when the set of messages $\mu_{i\setminus j}^{t}$ at time $t$ are zero and there is no observation also at time $t$, then the corresponding time-backward messages $\rho_{t \leftarrow}^{i\setminus j}(x_i^{t} )$ are also uniform. 
    Let us start from the time $T-1$, because by construction  $\mu_{i\setminus j}^{T}=0$, then using \eqref{eq:muij_rho} with the final time condition $\rho_{T \leftarrow}^{i \setminus j} ( x_i^T ) = p( {O}_i^{T} | x_i^T  )$ we obtain
    \begin{equation}
    \mu_{i\setminus j}^{t} =  \frac{1}{\mathcal{Z}_{i\setminus j}}\rho_{\rightarrow t}^{i\setminus j}\left(S\right)M_{T-1,SS}^{i\setminus j}\left(p\left( {O}_i^{T} | S  \right) - p\left( {O}_i^{T} | I  \right)\right).
    \end{equation}
    If no observation is provided on the final time, then $p\left(O_{i}^{T}| x_i^T\right)=1$ for $x_i^T=S,I$ and the numerator vanishes, that is $\mu_{i\setminus j}^{T-1}=0$. 
    Moreover, 
    \begin{subequations}
    \begin{align}
    \rho_{T-1\leftarrow}^{i\setminus j}\left(x_{i}^{T-1}\right)  & =  \sum_{x_{i}^{T}} M_{x_{i}^{T-1}x_{i}^{T}}^{i\setminus j} p\left(O_{i}^{T}| x_{i}^{T}\right)\\
    & =  M_{x_{i}^{T-1}S }^{i\setminus j} + M_{x_{i}^{T-1}I }^{i\setminus j},
    \end{align}
    \end{subequations}
    that is 
    \begin{align}
    \rho_{T-1\leftarrow}^{i\setminus j}\left(S\right)  &  =  g_{i\setminus j}^{T-1}p\left(O_{i}^{T-1}| S\right) + \left(1-g_{i\setminus j}^{T-1}\right) p\left(O_{i}^{T-1}| S\right) \nonumber\\
    &  = p\left(O_{i}^{T-1}| S\right),\\
    \rho_{T-1\leftarrow}^{i\setminus j}\left(I\right)  &  =   e^{\sum_{k\in\partial i\setminus j}\nu_{ik}^{T-1}\mu_{k\setminus i}^{T-1}}p\left(O_{i}^{T-1}| I\right)\nonumber\\
    & = p\left(O_{i}^{T-1}| I\right), 
    \end{align}
    meaning that $\rho_{T-1\leftarrow}^{i\setminus j}\left(S\right)=\rho_{T-1\leftarrow}^{i\setminus j}\left(I\right)$ if no observation is included at time $T-1$.
    In this way, the equality is guaranteed at time $T-1$ and one can proceed by induction. By assuming that, in the absence of observations at times larger than $t$, the equality
    is valid for time $t+1$, i.e. $\rho_{t+1\leftarrow}^{i\setminus j}\left(S\right)=\rho_{t+1\leftarrow}^{i\setminus j}\left(I\right)=\rho_{t+1\leftarrow}^{i\setminus j}$ for all edges $(i,j)$,
    then one obtains that $\mu_{i\setminus j}^t=0$. Computing the time-backward messages at time $t$,
    \begin{subequations}
    \begin{align}
    \rho_{t\leftarrow}^{i\setminus j}\left(x_{i}^{t}\right) & = \sum_{x_{i}^{t},x_{i}^{t+1}} M_{x_{i}^{t}x_{i}^{t+1}}^{i\setminus j} \rho_{t+1\leftarrow}^{i\setminus j} \left(x_{i}^{t+1}\right)\\
     & = \left(M_{x_{i}^{t}S}^{i\setminus j}+M_{x_{i}^{t}I}^{i\setminus j}\right) \rho_{t+1\leftarrow}^{i\setminus j},
    \end{align}
    \end{subequations}
    and using that all $\mu_{k\setminus i}^t$ vanish,
    \begin{align}
    \rho_{t\leftarrow}^{i\setminus j}\left(x_{i}^{t}\right) & \propto 
    \begin{cases} p\left(O_{i}^{t}| S\right) & \text{if $x_i^t=S$}, \\
    p\left(O_{i}^{t}| I\right) & \text{if $x_i^t=I$}.
    \end{cases}
    \end{align}
    By induction, this is true for every time, as long as no observation is included. Hence, it is possible to conclude that, in the absence of observations, the equations \eqref{eq:almostforward} for the cavity marginals $m_{i\setminus j}^{t}$ reduce to the more standard time-forward mean-field equations in Eqs.~\eqref{eq:mij_forward}.
    
    \section{Efficient implementation in the infection time representation}\label{app:efficientimplementation}
    
    As an alternative to the generic efficient formulation presented in Section~\ref{sec:efficientformulation} in terms of transfer-matrix formalism, the computational complexity of Eqs. \eqref{eq:noncausalMFm} and \eqref{eq:noncausalMFmu} can be reduced from exponential (in the temporal length $T$) to polynomial exploiting the non-recurrence of the SI dynamic process - in which only configurations of the type ${\bm x}_{i}=\left(0,\ldots,0,1,\ldots1\right)$ are allowed - by using a simpler representation in terms of infection times. An epidemic trajectory of the SI process can be parameterized by a unique set of integer variables $t_{i}$ (one for each node) representing the first time at which individual $i$ is infected, and taking values in
    $t_{i}\in\left\{ 0,\ldots,T+1\right\}$. The case $t_{i}=0$
    corresponds to individual $i$ being originally infected at the initial time, i.e. being a patient-zero of the epidemics. The other special case $t_{i}=T+1$ models the scenario where individual $i$ never gets infected during the dynamics (which formally corresponds to $t_{i}=+\infty$).  The trajectory ${\bm x}_{i}$ can be simply expressed as $x_{i}^{t}=S\Theta(t_i-1-t)+I\Theta(t-t_i)$,
    where $\Theta(x)$ is a Heaviside-step function, with the
    convention $\Theta(0)=1$.  After some algebra, we can rewrite Eqs.~\eqref{eq:noncausalMFm}, \eqref{eq:noncausalMFmu} and \eqref{eq:noncausalMFzij} as follows:
    \begin{widetext}
    \begin{subequations}
    \begin{align}
    m_{i\backslash j}^{t}&=\frac{1}{\mathcal{Z}_{i\setminus j}}\sum_{t_{i}=0}^{t}p\left({\bm O}_{i}| t_{i}\right)\gamma_{0}(t_i)\left(\prod_{r=0}^{t_{i}-2}\alpha_i^r e^{\sum_{k\in\partial i\setminus j}m_{k\setminus i}^{r}\nu_{ki}^{r}}\right)\left(1-\mathbb{I}_{1\leq t_{i}\leq T}\alpha_i^{t_i-1}e^{\sum_{k\in\partial i\setminus j}m_{k\setminus i}^{t_{i}-1}\nu_{ki}^{t_{i}-1}}\right)\prod_{s=t_{i}}^{T-1}e^{\sum_{k\in\partial i\setminus j}\nu_{ik}^{s}\mu_{k\setminus i}^{s}}\label{eq:mijt_infectimes} \\
    \mu_{i\backslash j}^{t} &= \frac{1}{\mathcal{Z}_{i\setminus j}}\sum_{t_{i}=t+2}^{T+1}p\left({\bm O}_{i}| t_{i}\right)\gamma_{0}(t_i)\left(\prod_{r=0}^{t_{i}-2}\alpha_i^r e^{\sum_{k\in\partial i\setminus j}m_{k\setminus i}^{r}\nu_{ki}^{r}}\right)\left(1-\mathbb{I}_{1\leq t_{i}\leq T}\alpha_i^{t_i-1}e^{\sum_{k\in\partial i\setminus j}m_{k\setminus i}^{t_{i}-1}\nu_{ki}^{t_{i}-1}}\right)\prod_{s=t_{i}}^{T-1}e^{\sum_{k\in\partial i\setminus j}\nu_{ik}^{s}\mu_{k\setminus i}^{s}}+\nonumber \\
     & \quad -\frac{\mathbb{I}_{0\leq t\leq T-1}}{\mathcal{Z}_{i\setminus j}}p\left({\bm O}_{i}| t+1\right) \gamma_{0}(t+1)\left(\prod_{r=0}^{t} \alpha_i^r e^{\sum_{k\in\partial i\setminus j}m_{k\setminus i}^{r}\nu_{ki}^{r}}\right)\prod_{s=t+1}^{T-1}e^{\sum_{k\in\partial i\setminus j}\nu_{ik}^{s}\mu_{k\setminus i}^{s}}\label{eq:muijt_infectimes}\\
     \mathcal{Z}_{i\setminus j}&=\sum_{t_{i}=0}^{T+1}p\left({\bm O}_{i}| t_{i}\right)\gamma_{0}(t_i)\left(\prod_{r=0}^{t_{i}-2}\alpha_i^r e^{\sum_{k\in\partial i\setminus j}m_{k\setminus i}^{r}\nu_{ki}^{r}}\right)\left(1-\mathbb{I}_{1\leq t_{i}\leq T}\alpha_i^{t_i-1}e^{\sum_{k\in\partial i\setminus j}m_{k\setminus i}^{t_{i}-1}\nu_{ki}^{t_{i}-1}}\right)\prod_{s=t_{i}}^{T-1}e^{\sum_{k\in\partial i\setminus j}\nu_{ik}^{s}\mu_{k\setminus i}^{s}}\label{eq:zij_infectimes}
    \end{align}
    \end{subequations}
    \end{widetext}
    where the function $\gamma_{0}(t_i)$ is related to the probability of node $i$ being a patient zero, namely
    \begin{equation}
    \gamma_{0}\left(t_i\right)=\begin{cases}
    p_0(x_i^0=I) & t_i=0\\
    1-p_0(x_i^0=I) & t_i>0.
    \end{cases}
    \end{equation}
    We have introduced the indicator function $\mathbb{I}_x$ which is equal to one when its argument $x$ is true, and zero otherwise.
    Notice how all the summations and products w.r.t. the infection times are now linear in $T$. Analogously, the likelihood term for each observations on node $i$ (Eq.~\eqref{eq:likelihood_singleobs}) can be rewritten under this representation as
    \begin{align}
    p\left(O_{i}^t=S|t_{i}\right) & =
    (1-\fpr)\Theta(t_{i}-t-1)+\fnr\Theta(t-t_{i}),\\
    p\left(O_{i}^t=I|t_{i}\right) & = \fpr\Theta(t_{i}-t-1)+(1-\fnr)\Theta(t-t_{i}),
    \end{align}
    where $p({\bm O}_i|t_i)=\prod_{t=0}^{T}p\left(O_{i}^t|t_{i}\right)$.
    Analogous expressions w.r.t. Eqs.~\eqref{eq:mijt_infectimes} and \eqref{eq:zij_infectimes} can be derived for the single-node marginal $m_i^t$ and its normalization $\mathcal{Z}_i$.
    An efficient computational scheme can be attained by updating all the cavity quantities of a fixed node at once, and then performing a random shuffling on the order in which nodes are updated. Intuitively, the speed-up induced by this protocol is that the forward and backward contribution to each cavity message - say, on link $(i,j)$ can be computed by removing the corresponding link contribution from the \textit{site} term $i$. 
    In order to clarify this point, let us define the following four quantities, for each node $i$:
    \begin{align}
    \mathcal{R}_{i}^{\rightarrow t} & =\log \alpha_{i}^{t} +  \sum_{k\in\partial i}m_{k\setminus i}^{t}\nu_{ki}^{t},\label{eq:Ri_forward}\\
    \mathcal{R}_{i}^{t \leftarrow} & =\sum_{k\in\partial i}\nu_{ik}^{t}\mu_{k\setminus i}^{t},\label{eq:Ri_backward}\\
    K_{i}^{\rightarrow t}& =\sum_{r=0}^{t-2}\mathcal{R}_{i}^{\rightarrow r},\label{eq:Ki_forward}\\ 
    K_{i}^{t \leftarrow}& =\sum_{s=t}^{T-1}\mathcal{R}_{i}^{s \leftarrow}.\label{eq:Ki_backward}
    \end{align}
    Analogous definitions for the cavity quantities hold for each edge $(i, j)$, just by letting the above summations run over all the neighbors of node $i$ but $j$.  Eqs.~\eqref{eq:Ri_forward} and \eqref{eq:Ri_backward} have also a physical interpretation. For instance, $\mathcal{R}_{i}^{\rightarrow t}$ is a mean-field approximation for the log-probability of node $i$ not being infected at time $t$ by none of its neighbors.
    The above definitions Eqs.~\eqref{eq:Ri_forward}-\eqref{eq:Ki_backward} allow one to write the cavity equations Eqs.~\eqref{eq:mijt_infectimes}-\eqref{eq:zij_infectimes} in a more compact form:
    \begin{align}
    m_{i\backslash j}^{t} & = \frac{1}{\mathcal{Z}_{i\setminus j}}\sum_{t_{i}=0}^{t}p\left({\bm O}_{i}| t_{i}\right)\gamma_{0}(t_i)e^{K_{i\backslash j}^{\rightarrow t_i}}\theta_{i\setminus j}^{\rightarrow t_i-1}e^{K_{i\backslash j}^{t_i \leftarrow}},\\
    \mu_{i\backslash j}^{t} & = \frac{1}{\mathcal{Z}_{i\setminus j}}\Bigg(\sum_{t_{i}=t+2}^{T+1}p\left({\bm O}_{i}| t_{i}\right)\gamma_{0}(t_i)e^{K_{i\backslash j}^{\rightarrow t_i}}\theta_{i\setminus j}^{\rightarrow t_i-1}e^{K_{i\backslash j}^{t_i \leftarrow}}\nonumber \\
     & \; - p\left({\bm O}_{i} | t+1\right)\gamma_{0}(t+1)\mathbb{I}_{0\leq t\leq T-1}e^{K_{i\backslash j}^{\rightarrow t+2}+K_{i\backslash j}^{t+1 \leftarrow}}\Bigg),\\
    \mathcal{Z}_{i\setminus j} & = \sum_{t_{i}=0}^{T+1}p\left({\bm O}_{i}| t_{i}\right)\gamma_{0}(t_i)e^{K_{i\backslash j}^{\rightarrow t_i}}\theta_{i\setminus j}^{\rightarrow t_i-1}e^{K_{i\backslash j}^{t_i \leftarrow}},
    \end{align}
    where we have introduced $\theta_{i\setminus j}^{\rightarrow t}=1-\mathbb{I}_{1\leq t+1\leq T}e^{\mathcal{R}_{i\setminus j}^{\rightarrow t}}$.
    
    At fixed $i$, the quantities $\mathcal{R}_{i\backslash j}^{\rightarrow t}$, $\mathcal{R}_{i\backslash j}^{t \leftarrow}$, $K_{i\backslash j}^{\rightarrow t}$, $K_{i\backslash j}^{t \leftarrow}$ can be computed for each cavity by removing just one link contribution (i.e. the one corresponding to the link removed in that specific cavity graph), without further $O\left( \left| \partial i \right | \right) $ computations, namely
    \begin{align}
    \mathcal{R}_{i\backslash j}^{\rightarrow t} & =\mathcal{R}_{i}^{\rightarrow t}-m_{j\backslash i}^{t}\nu_{ji}^{t},\\ 
    \mathcal{R}_{i\backslash j}^{t \leftarrow} & =\mathcal{R}_{i}^{t \leftarrow}-\nu_{ij}^{t}\mu_{j\backslash i}^{t},\\
    K_{i\backslash j}^{\rightarrow t} & = K_{i}^{\rightarrow t}-\sum_{r=0}^{t-2}m_{j\backslash i}^{r}\nu_{ji}^{r},\\ 
    K_{i\backslash j}^{t \leftarrow} & = K_{i}^{t \leftarrow} - \sum_{s=t}^{T-1}\nu_{ij}^{s}\mu_{j\backslash i}^{s},
    \end{align}
    for any $j\in \partial i$.
    Furthermore, the computation of $K_i^{\rightarrow t}$ and $K_i^{t\leftarrow}$ can be done recursively, in a forward (resp. backward)
    direction w.r.t. time,
    i.e. by exploiting $K_{i}^{\rightarrow t} = K_{i}^{\rightarrow t-1} +\mathcal{R}_{i}^{\rightarrow t-2}$ and 
    $K_{i}^{t \leftarrow} = K_{i}^{t+1 \leftarrow}+\mathcal{R}_{i}^{t\leftarrow}$.
    Clearly, equivalent relations hold for the cavity quantities $K_{i\backslash j}^{\rightarrow t}$ and $K_{i\backslash j}^{t \leftarrow}$.
    Using all the above schemes, the overall computational cost to perform a single update of all the cavity quantities for a node $i$ scales as $O\left(\left|\partial i\right| T \right)$.
    A further advantage of such a computational scheme is that the update of all the cavities for one node can be performed in parallel, a particularly convenient choice especially when dealing with dense graphs.
    The convergence criterion can be defined either w.r.t. the cavity messages $\{ m_{i\backslash j}^t\}$ and/or their conjugates $\{\mu_{i\backslash j}^t \}$, or eventually w.r.t. the single-site marginals $\{m_{i}^t\}$: the latters can be computed as
    \begin{equation}
    m_{i}^{t}=\frac{\sum_{t_{i}=0}^{t}p\left({\bm O}_{i}| t_{i}\right)\gamma_{0}(t_i)e^{K_{i}^{\rightarrow}(t_i)}\theta_{i}^{t_i-1}e^{K_{i}^{\leftarrow}(t_i)}}{\sum_{t_{i}=0}^{T+1}p\left({\bm O}_{i}| t_{i}\right) \gamma_{0}(t_i)e^{K_{i}^{\rightarrow}(t_i)}\theta_{i}^{t_i-1}e^{K_{i}^{\leftarrow}(t_i)}}\label{eq:mi_t_short}
    \end{equation}
    where the normalization is explicitly shown at the denominator, and $\theta_{i}^{t}$ is the straightforward extension of $\theta_{i\setminus j}^{t}$ to the single node case. This expression is equivalent to Eq.~\eqref{eq:marginals_mi_main} of the main text but rewritten using the infection-time representation just discussed.
    
    \section{Derivation of SCDC on the SIR model}\label{app:SIRmodel}
    
    In a Susceptible-Infected-Recovered (SIR) model, the possible individual states are $x_i^t \in \{S, I, R \}$ and the transition probabilities between states are given by
    \begin{align}
        W_i(x_i^{t+1}=S|{\bm x}^t) & = \delta_{x_i^t,S}\alpha_i^t e^{h_i^t}\\
        W_i(x_i^{t+1}=I|{\bm x}^t) & = \delta_{x_i^t,I}(1-r_i^t) + \delta_{x_i^t,S} \left[1-\alpha_i^t e^{h_i^t}\right]\\
        W_i(x_i^{t+1}=R|{\bm x}^t) & = \delta_{x_i^t,R} + \delta_{x_i^t,I}r_i^t 
    \end{align}
    where $r_i^t$ is the recovery probability of individual $i$ at time $t$ and $h_i^t$ is the usual local field, defined as $h_{i}^{t}=\sum_{j=1}^N \nu_{ji}^{t}\delta_{x_{j}^{t},I}$, where $\nu_{ji}^{t}=\log\left(1-\lambda_{ji}^{t}\right)$, such that $e^{h_{i}^{t}}=\prod_{j=1}^N(1- \lambda_{ji}^{t}\delta_{x_{j}^{t},I})$ is the probability of not being infected by the neighbors.
    The dynamical partition function of the system can be computed following the same steps of the derivation for the SI model, i.e. introducing the local fields $h_i^t$ through the integral representation of a Dirac delta function, and integrating over them at all them. After some calculations, we end up with the following expression for the dynamical partition function of the observation reweighted SIR dynamics
    \begin{widetext}
    \begin{align}
        \mathcal{Z}(\boldsymbol{\mathcal{O}}) & = \sum_{{\bm X}}\int \mathcal{D}\hat{{\bm H}}\prod_i p(x_i^0)p_0({\bm O}_i|{\bm x}_i)\prod_t \left\{\delta(\hat{h}_i^t)\left[\delta_{x_i^{t+1},I}\left(\delta_{x_i^t,I}(1-r_i^t)+\delta_{x_i^t,S} \right)+\delta_{x_i^{t+1},R}\left(\delta_{x_i^t,R}+\delta_{x_i^t,I}r_i^t \right) \right]+ \right. \nonumber \\
        &\qquad \left.+\delta(\hat{h}_i^t-{\rm i})\alpha_i^t \left[\delta_{x_i^{t+1},S}\delta_{x_i^t,S}-\delta_{x_i^{t+1},I}\delta_{x_i^t,S} \right] \right\}  \prod_{j>i}e^{\delta_{x_j^t,I}\nu_{ji}^t(-i\hat{h}_i^t)+\delta_{x_i^t,I}\nu_{ij}^t(-i\hat{h}_j^t)}
    \end{align}
    which can in turn be interpreted as a graphical model in a similar way to what was done for the SI model. The dynamic cavity equations for the SIR model with observations have the same form of the SI model (Eq.~\eqref{eq:dyncav_fourier}), where the only difference is in the definition of the sum over the trajectories, which now runs over $\{S,I,R\}$, and in the definition of the local non-interacting action
    \begin{align}
    S_i^0 & =\sum_t\log\left\{\delta(\hat{h}_i^t)\left[\delta_{x_i^{t+1},I}\left(\delta_{x_i^t,I}(1-r_i^t)+\delta_{x_i^t,S} \right)+\delta_{x_i^{t+1},R}\left(\delta_{x_i^t,R}+\delta_{x_i^t,I}r_i^t \right) \right]+ \right.\nonumber \\
    &\qquad \left.+\delta(\hat{h}_i^t-{\rm i})\alpha_i^t\left[\delta_{x_i^{t+1},S}\delta_{x_i^t,S}-\delta_{x_i^{t+1},I}\delta_{x_i^t,S} \right]\right\}
    \end{align}
    The expansion for small infection rates can be carried out similarly to the SI model, leading to the same expression for the cavity averages Eqs.~\eqref{eq:noncausalMFm} and \eqref{eq:noncausalMFmu}, where the two term $S_{i\setminus j}^m$ and $S_{i\setminus j}^\mu$ are now defined as
    \begin{align}
    S_{i\setminus j}^m &= \sum_t  \log\left\{\delta_{x_i^{t+1},I}\left(\delta_{x_i^t,I}(1-r_i^t)+\delta_{x_i^t,S} \right)+\delta_{x_i^{t+1},R}\left(\delta_{x_i^t,R}+\delta_{x_i^t,I}r_i^t \right) \right. \nonumber \\
    &\qquad \left.+\alpha_i^t\left(\delta_{x_i^{t+1},S}\delta_{x_i^t,S}-\delta_{x_i^{t+1},I}\delta_{x_i^t,S} \right) e^{\sum_{k \in \partial i \setminus j}m_{k\setminus i}^t\nu_{ki}^t}\right\} +\sum_t \sum_{k \in \partial i \setminus j} \delta_{x_i^t,I}\nu_{ik}^t\mu_{k\setminus i}^t,\\
    S_{i\setminus j}^\mu & = \sum_t \log\left[\alpha_i^t\left(\delta_{x_i^{t+1},S}\delta_{x_i^t,S}-\delta_{x_i^{t+1},I}\delta_{x_i^t,S} \right)\right] + \sum_t \sum_{k \in \partial i \setminus j} \left( m_{k\setminus i}^t\nu_{ki}^t+ \delta_{x_i^t,I}\nu_{ik}^t\mu_{k\setminus i}^t\right).
    \end{align}
    \end{widetext}
    The normalization function is defined as usual by Eq.~\eqref{eq:noncausalMFzij}, and ensures normalization of the cavity averages $m_{i\setminus j}^t$, which are therefore interpreted as the probability of node $i$ of being infected at time $t$ when node $j$ is removed from the contact graph. Even with the new definitions of $S_{i\setminus j}^m$ and $S_{i\setminus j}^\mu$, we can use the efficient formulation Eq.~\eqref{eq:partition_function} in terms of transfer matrices $M_{x_i^t,x_i^{t+1}}$, with the only difference in their expression. The transition matrices are now 3x3, and defined by Eq.~\eqref{eq:transferMatrixSIR}.

    \section{Example of normalization issue for leaves of the contact graph}\label{app:leaves}
    
    The small coupling expansion requires to assume the normalization $\mathcal{Z}_{i\setminus j}$ in \eqref{eq:noncausalMFzij}, which sums over all the possible trajectories of node $i$ assuming $\boldsymbol{s}_{i} = 0$ at every time. The method thus considers all the trajectories in which the cavity node $j$ is always susceptible, and therefore cannot infect node $i$. In particular, there are situations, such as the one considered in the example below, in which the normalization vanishes, meaning that it is not possible to explain an observed trajectory within the standard SI model. While this could seem pathological, it is worth stressing that the assumption done is necessary to obtain a message-passing algorithm that is independent of the trajectory of node $j$, a crucial condition to perform the expansion on which the present method is based. It is however possible to ensure that every trajectory of a node $i$ remains feasible, the normalization constant being finite, by slightly modifying the epidemic model introducing a small self-infection probability. In addition to fix the normalization issue, a small value of self-infection probability does not deteriorate the predictive power of the method. 
    
    To better illustrate this problem, we consider a leaf node $i$ and its unique neighbor $j$. In the cavity graph corresponding to the message $c_{i\setminus j}({\bm x}_i, {\bm s}_i)$, node $i$ will appear as an isolated node. As a consequence, it is expected that the approximation behind the SCDC equations cannot explain, within the cavity graph, an infection actually transmitted from node $j$ to node $i$. Indeed, because of the absence of further neighbors, the normalization term reads, after integration of the conjugate field ${\bm h}_i$,
    \begin{align}
    \mathcal{Z}_{i\setminus j} & =\sum_{{\bm x}_{i}}p_0(x_{i}^{0})p({\bm O}_i|{\bm x}_i) \nonumber \\
    &\quad \times \prod_t\left[\alpha_i^t(\delta_{x_{i}^{t+1},x_{i}^{t}}-\delta_{x_{i}^{t+1},I})+\delta_{x_i^{t+1},I}\right],
    \end{align}
    showing that an infection can only be explained by a self-infection event. When $\varepsilon_i^t =0$ (i.e. $\alpha_i^t=1$, no self-infections possible), the cavity message admits trajectories for which node $i$ is always susceptible or infected. When a perfect observation (i.e. $\fpr=\fnr=0$) imply an infection event at some $t\neq 0$, the normalization vanishes, indicating an inconsistency in the model. This is prevented by the existence of a finite self-infection probability. Since it is recommended to operate in the limit of a vanishing self-infection, in the present case it is possible to analytically verify  the limiting behavior for the cavity messages $m_{i\setminus j}^t$ and $\mu_{i\setminus j}^t$.   
    
    As an example, we suppose that the leaf $i$ is observed to be susceptible at time $t_S$ and then is observed to be infected at time $t_I > t_S$. We consider a uniform self-infection probability $\varepsilon_i^t = \varepsilon$ ($\alpha_i^t=\alpha=1-\varepsilon$) for any time $t$ and any node $i$, and a uniform prior probability $p_0\left( x_i^0 = S\right) = 1-\gamma$, $p_0\left( x_i^0 = I\right) = \gamma$. The forward  messages are
    \begin{align}
    \rho_{\rightarrow t}^{i\setminus j}\left(S\right) & =
    \begin{cases} (1-\gamma)\alpha^t & \text{if $t \leq t_I$}, \\
    0 & \text{if $t> t_I$},
    \end{cases}\\
    \rho_{\rightarrow t}^{i\setminus j}\left(I\right) & =
    \begin{cases} \gamma + (1-\alpha)(1-\gamma)\sum_{l = 0}^{t-1}\alpha^l & \text{if $t \leq t_S$}, \\
    (1-\alpha)(1-\gamma)\sum_{l = t_S}^{t-1}\alpha^l & \text{if $t_S < t \leq t_I$}, \\
    (1-\alpha)(1-\gamma)\sum_{l = t_S}^{t_I-1}\alpha^l & \text{if $t>t_I$},\\
    \end{cases}
    \end{align}
    and the backward messages 
    \begin{align}
    \rho_{t\leftarrow}^{i\setminus j}\left(S\right) & =
    \begin{cases} (1-\alpha)\alpha^{t_S-t}\sum_{l=0}^{t_I-1-t_S}\alpha^l & \text{if $t \leq t_S$}, \\
    (1-\alpha)\sum_{l=0}^{t_I-1-t}\alpha^l & \text{if $t_S < t \leq t_I$},\\
    1 & \text{if $t>t_I$}
    \end{cases}\\
    \rho_{t\leftarrow}^{i\setminus j}\left(I\right) & =
    \begin{cases} 0 & \text{if $t \leq
    t_S$}, \\
    1 & \text{if $t > t_S$}.
    \end{cases}
    \end{align}
    
    The normalization factor, taking into account the observations, is given by
    \begin{equation}
        \mathcal{Z}_{i\setminus j} = (1-\alpha)(1-\gamma)\sum_{l = t_S}^{t_I-1}\alpha^l, 
    \end{equation}
    so that the cavity marginal is given by 
    \begin{align}
        m_{i\setminus j}^t = 
        \begin{cases}
            0 & \text{if $t\leq t_S$},\\
            \frac{(1-\alpha)(1-\gamma)\sum_{l = t_S}^{t-1}\alpha^l}{(1-\alpha)(1-\gamma)\sum_{l = t_S}^{t_I-1}\alpha^l} & \text{if $t_S < t \leq t_I$}, \\
            1 & \text{if $t>t_I$}.\\
        \end{cases}
    \end{align}
    In the limit of vanishing self-infection $\varepsilon \to 0$ ($\alpha\to 1$), the cavity marginal takes the simple expression 
    \begin{align}
        m_{i\setminus j}^t = 
        \begin{cases}
            0 & \text{if $t\leq t_S$},\\
            \frac{t-t_S}{t_I-t_S} & \text{if $t_S < t \leq t_I$}, \\
            1 & \text{if $t>t_I$},\\
        \end{cases}
    \end{align}
    which gives a reasonable probability profile for the node $i$ to be infected in the absence of node $i$. It is worth stressing that this is not the full marginal $m_i^t$, which also depends on the messages coming from $j$ to $i$.  
    The cavity field instead diverges for times $t$ between the two observation times
    \begin{align}
        \mu_{i\setminus j}^t = 
        \begin{cases}
            1 & \text{if $t\leq t_S$},\\
            -\infty & \text{if $t_S < t \leq t_I$}, \\
            0 & \text{if $t>t_I$},\\
        \end{cases}
    \end{align}
    which is a clear consequence of having a vanishing normalization factor in the limit of zero self-infection. The divergence of the cavity field is thus the very non-physical effect of the inconsistency already discussed. 
    In practice, in  order to avoid divergences triggered by some peculiar combinations of observations, we then implement the algorithm using a cutoff $\mu_{c}<0$ on the values of $\mu_{i\setminus j}$, such that the update rule \eqref{eq:muij_rho} is implemented as follows  
    \begin{equation}
        \mu_{i\setminus j}^t = \max\left\{\mu_{c}, \frac{\rho_{\rightarrow t}^{i\setminus j}(S)M_{SS}^{i\setminus j}\left(\rho_{t+1\leftarrow}^{i\setminus j}\left(S\right) - \rho_{t+1\leftarrow}^{i\setminus j}\left(I\right)\right)}{\rho_{\rightarrow t}^{i\setminus j}\left(S\right)\rho_{t\leftarrow}^{i\setminus j}\left(S\right) + \rho_{\rightarrow t}^{i\setminus j}\left(I\right)\rho_{t\leftarrow}^{i\setminus j}\left(I\right)} \right\}.
    \end{equation}

\end{document}